\documentclass[aps,jcp,preprint,superscriptaddress]{revtex4-1}

\usepackage{url}
\usepackage{amssymb}
\usepackage{mathtools}
\usepackage{graphicx}
\usepackage{xcolor}
\usepackage{setspace}

\newcommand{\dint}[1]{\mathrm{d}{#1}}

\newcommand{\dpart}[2]{\frac{\mathrm{\partial}{#1}}{\mathrm{\partial}{#2}}}

\newcommand{\kBT}{k_\mathrm{B} T}

\newcommand{\Pe}{\mathrm{Pe}}
\newcommand{\Bi}{\mathrm{Bi}}

\newcommand{\Ca}{\mathrm{Ca}}
\newcommand{\Cn}{\mathrm{Cn}}
\newcommand{\G}{\mathrm{G}}

\date{\today}
 
\begin{document}
 \title{Structuring in Thin Films during Meniscus-Guided Deposition}
	\author{René de Bruijn}
	\email{r.a.j.d.bruijn@tue.nl}
	\affiliation{Department of Applied Physics and Science Education, Eindhoven University of Technology, P.O. Box 513, 5600 MB Eindhoven, The Netherlands}
	\affiliation{Institute for Complex Molecular Systems, Eindhoven University of Technology, P.O. Box 513, 5600 MB Eindhoven, The Netherlands}
 	\author{Anton A. Darhuber}
	\affiliation{Department of Applied Physics and Science Education, Eindhoven University of Technology, P.O. Box 513, 5600 MB Eindhoven, The Netherlands}
 	\author{Jasper J. Michels}
	\affiliation{Max Planck Institute for Polymer Research, Mainz, Germany}
  	\author{Paul van der Schoot}
	\affiliation{Department of Applied Physics and Science Education, Eindhoven University of Technology, P.O. Box 513, 5600 MB Eindhoven, The Netherlands}
\begin{abstract}
We study theoretically the evaporation-driven phase separation of a binary fluid mixture in a thin film deposited on a moving substrate, as occurs in meniscus-guided deposition for solution-processed materials. Our focus is on rapid substrate motion during, where phase separation takes place far removed from the coating device under conditions where the mixture is essentially stationary with respect to the substrate. We account for the hydrodynamic transport of the mixture within the lubrication approximation. In the early stages of demixing, diffusive and evaporative mass transport predominates, consistent with earlier studies on evaporation-driven spinodal decomposition. By contrast, in the late-stage coarsening of the demixing process, the interplay of solvent evaporation, diffusive, and hydrodynamic mass transport results in a number of distinct coarsening mechanisms. The effective coarsening rate is dictated by the (momentarily) dominant mass transport mechanism and therefore depends on the material properties, evaporation rate and time: slow solvent evaporation results in initially diffusive coarsening that for sufficiently strong hydrodynamic transport transitions to hydrodynamic coarsening, whereas rapid solvent evaporation can preempt and suppress either or both hydrodynamic and diffusive coarsening. We identify a novel hydrodynamic coarsening regime for off-critical mixtures, arising from the interaction of the interfaces between solute-rich and solute-poor regions in the film with the solution-gas interface. This interaction induces directional motion of solute-rich droplets along gradients in the film thickness, from regions where the film is relatively thick to where it is thinner. The solute-rich domains subsequently accumulate and coalesce in the thinner regions, enhancing domain growth.
\end{abstract}
\maketitle

\section{Introduction}
Solution-processed thin films are an essential component in the production of organic electronics with applications ranging from organic photovoltaics to sensors, transistors and many more~\cite{Janssen2019,Mei2013,Janasz2022OrganicSensors}. The films are commonly manufactured by dissolving the constituents in a solution containing one or more volatile solvents, and subsequently deposited onto a substrate where the solvent is removed by drying~\cite{Chen2020UnderstandingFabrications}. In the course of the drying of the (liquid) film a very complex microscopic morphology emerges~\cite{Bornside1989SpinModel,Diao2014MorphologyFilms}, which typically forms via phase separation, crystallization or a combination of both~\cite{Diao2014MorphologyFilms,Chen2020UnderstandingFabrications}. This morphology is believed to be crucial for the (efficient) functioning of the devices~\cite{Chen2020UnderstandingFabrications,Gu2016ComparisonCoating,Wang2021ThePerformances} and therefore the ability to control the emergent morphology is of paramount importance for the rational design of organic electronic devices~\cite{Chen2020UnderstandingFabrications,Schaefer2015StructuringEvaporation,Peng2023ASemiconductors}. 

One of the key factors affecting the final dry film morphology are the processing settings, which are often specific to any particular deposition technique~\cite{Schaefer2015StructuringEvaporation,Negi2018SimulatingInvestigation,Yildiz2022OptimizedCoating,Franeker2015cosolvents,Franeker2015spincoating,vanFraneker2015PolymerFormation}. For instance, in spin coating the rate of solvent evaporation, controlled by the so-called spin speed, crucially governs the morphology~\cite{Schaefer2015StructuringEvaporation,Schaefer2016StructuringEvaporation}, impacting both the initial demixing dynamics~\cite{Schaefer2015StructuringEvaporation,Schaefer2016StructuringEvaporation,deBruijn2024TransientEvaporation} and late-stage coarsening~\cite{Schaefer2016StructuringEvaporation,Negi2018SimulatingInvestigation}. Another frequently used family of deposition techniques is meniscus-guided deposition, where the solution is deposited from a stationary dispensing unit onto a moving substrate~\cite{Diao2014MorphologyFilms}. For this technique the substrate velocity and hydrodynamic transport processes in the film and meniscus that are present due to the directional motion of the substrate become important control variables also~\cite{deBruijn2024PeriodicDeposition,Yildiz2022OptimizedCoating,Michels2021PredictiveCoating,Rogowski2011SolutionProperties}. In our recent study of the meniscus-guided deposition of a binary phase-separating mixture, we show that these factors can significantly affect the morphology but only if the substrate moves sufficiently slowly. Indeed, the substrate should move more slowly than the growth rate of the demixed structures~\cite{deBruijn2024PeriodicDeposition}. At high velocities that typically are within the so-called Landau-Levich regime, the solution dries significantly only far removed from the dispensing unit under conditions that arguably resemble those during spin coating~\cite{Schaefer2015StructuringEvaporation,Negi2018SimulatingInvestigation,Schaefer2016StructuringEvaporation,Ronsin2022FormationSimulations,Franeker2015spincoating}. Under these conditions, hydrodynamic transport processes related to the deposition technique itself should not impact the demixed morphology.

Even if hydrodynamics related to the deposition process is negligible, hydrodynamic transport processes due to solvent evaporation, the evolution of the free solution-gas surface and demixing do remain important irrespective of the deposition technique. For non-volatile mixtures the impact of hydrodynamics on the morphological evolution of a demixing solution, whether in bulk or confined between parallel plates, has been studied extensively  theoretically~\cite{Siggia1979LateMixtures,Tanaka1996CoarseningMixtures,Bray2002TheoryKinetics,Zoumpouli2016HydrodynamicSolutions}, numerically~\cite{Tanaka1996CoarseningMixtures,Chen1997SurfaceEffects,Tanaka2001InterplayHydrodynamics,Zoumpouli2016HydrodynamicSolutions} and by means of experiments~\cite{Tanaka2001InterplayHydrodynamics,Bouttes2015HydrodynamicMelts,Sung1996DimensionalFilms,Song1995CoarseningSolutions,Haas1997Two-dimensionalFilms}. In general, it seems that hydrodynamics is relevant only in the coarsening stage of the demixed morphology. As is often observed in experiments and well understood theoretically, coarsening is characterized by a characteristic feature size $\langle L \rangle$ that adheres to a power law relation $\langle L \rangle \propto t^\alpha$ with $\alpha$ the coarsening exponent that depends on the dominant mass transport mechanism~\cite{Tanaka1996CoarseningMixtures,Siggia1979LateMixtures,Bray2002TheoryKinetics}. For connected or bicontinuous morphologies, coarsening transitions have been predicted and observed, from diffusive coarsening with an exponent of either $\alpha = 1/3$ or $1/4$, depending on the diffusive mobilities of the solute and the solvent molecules, to viscous coarsening with an exponent of $\alpha = 1/2$ in two dimensions and $\alpha = 1$ in three dimensions~\cite{Siggia1979LateMixtures,Bray2002TheoryKinetics,Lifshitz1961TheSolutions,Wagner1961TheorieOstwaldReifung}. A second transition exists from viscous to inertial coarsening with for the latter an exponent $\alpha = 2/3$ in both two and three dimensions~\cite{Siggia1979LateMixtures,Bray2002TheoryKinetics}. These coarsening regimes are, however, absent if the morphology is disconnected, as is the case for off-critical mixtures wherein one of the phases forms droplets. Short-ranged hydrodynamic interactions between the droplets still operate, which tend to facilitate their coalescence either via attractive Marangoni-like interactions~\cite{Shimizu2015AMixtures} or via a cascade of coalescence events, because coalescing droplets result in motion of the surrounding fluid~\cite{Tanaka1996CoarseningMixtures,Chen1997SurfaceEffects,Tanaka2001InterplayHydrodynamics}. Such short-ranged interactions give rise to a coarsening exponent of $\alpha = 1/3$ and is therefore in this sense often indistinguishable from diffusive coarsening~\cite{Tanaka1996CoarseningMixtures,Shimizu2015AMixtures}. 

In contrast to the studies on non-volatile mixtures, many theoretical and numerical studies on volatile solutions have neglected both the phase-separation hydrodynamics and the hydrodynamics caused by solvent evaporation itself~\cite{Schaefer2015StructuringEvaporation,Negi2018SimulatingInvestigation,Schaefer2016StructuringEvaporation}. Only recently attention has shifted to include these transport processes in order to better mimic the conditions and transport processes taking place during the solution processing of thin (polymeric) films~\cite{Zoumpouli2016HydrodynamicSolutions,Ronsin2022FormationSimulations,Ronsin2022PhaseFieldFilms,Cummings2018ModelingMixtures}. These studies are, however, limited to the case of a stationary film, whereas films are frequently fabricated via deposition on a moving substrate. Moreover, we are not aware of any systematic studies on the effect of hydrodynamics on the phase-separation kinetics in a volatile thin film, during either the early-stage demixing or the late-stage coarsening. 
In this work, we investigate by means of numerical calculations the effect of hydrodynamic transport processes in a binary mixture confined to a thin film, undergoing evaporation-driven spinodal decomposition on a moving substrate in a meniscus-guided deposition setup. We focus in particular on the limit of rapid substrate motion in the so-called Landau-Levich regime, where phase separation occurs far from the meniscus, under conditions where the mixture is for all intends and purposes stationary with respect to the substrate and the solution-gas interface is parallel to the substrate. Following our earlier work in the slow-coating evaporative regime~\cite{deBruijn2024PeriodicDeposition}, we here also treat the (hydrodynamic) transport processes within a height-averaged approximation, suppressing stratification in the film~\cite{deBruijn2024PeriodicDeposition, Thiele2016GradientSurfactant,Clarke2005TowardMixtures,Naraigh2007DynamicalFilms,Naraigh2010NonlinearFilms}. This is reasonable for films that are sufficiently thin, in absence of preferential interactions of the components in solution with the substrate or solution-gas interface and for sufficiently slow solvent evaporation~\cite{Clarke2005TowardMixtures,Naraigh2010NonlinearFilms,Thiele2016GradientSurfactant,Larsson2022QuantitativeFilms}. 

\begin{figure}
    \centering
    \includegraphics[width=0.49\columnwidth]{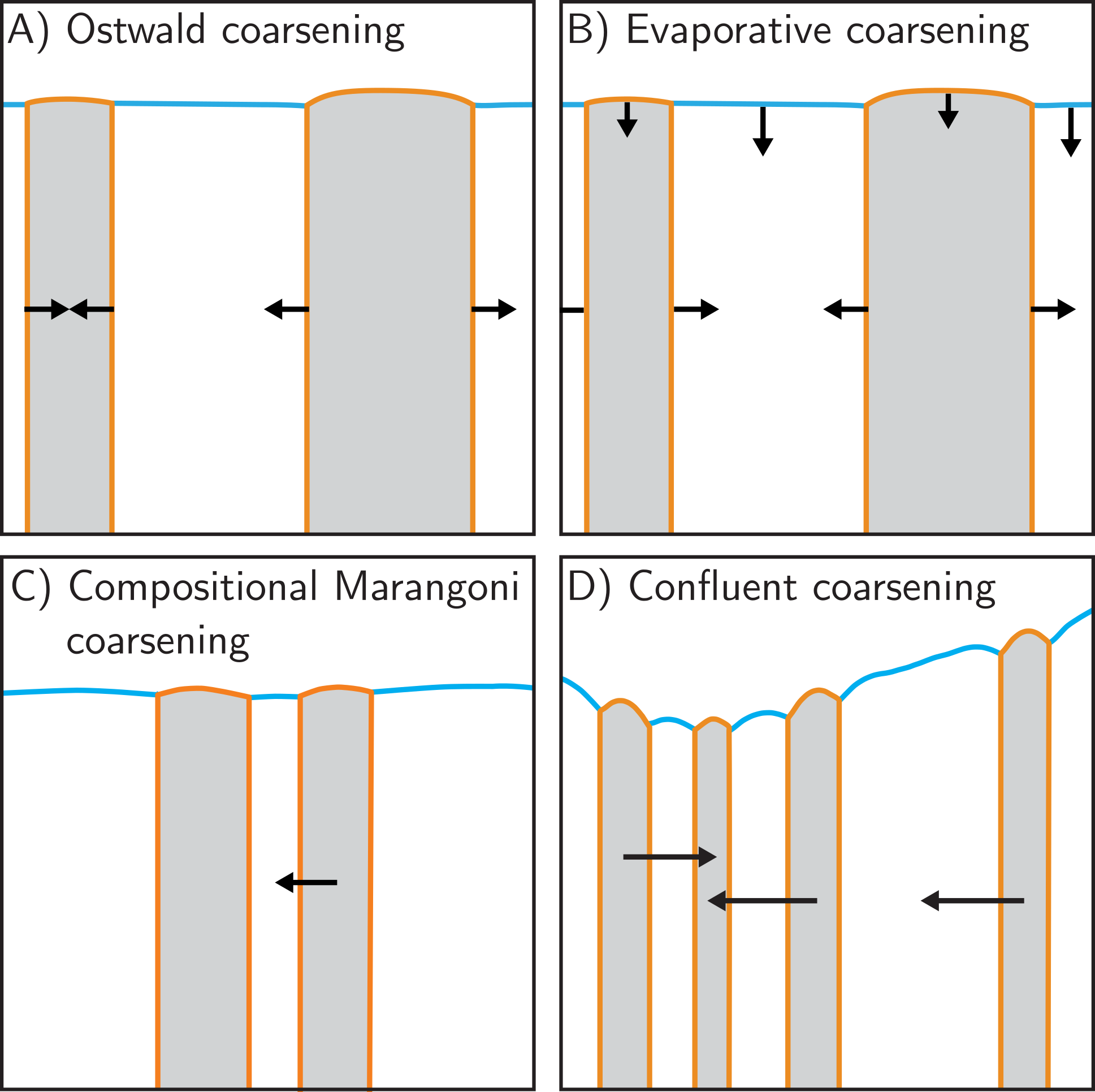}
    \caption{A schematic representation of the four main coarsening mechanisms that we find in our numerical calculations. Shown is a side view of the phase separating film (not to scale). The solute-rich domains are shown in gray with the fluid phase boundaries indicated in orange. The blue line represents the free solution-gas interface. In panel A we show the classical Ostwald-ripening that originates from differences in the Laplace pressure between small and large domains. In panel B we depict evaporative coarsening where the decreasing thickness of the thin film results in the lateral redistribution of solute mass. Panel C illustrates the short-ranged attractive hydrodynamic interactions known as the compositional Marangoni effect~\cite{Shimizu2015AMixtures}, which originates from gradients in the solute-solvent surface tension that itself find their origin in the Ostwaldian mass transport from small to large droplets. In panel D we highlight what we refer to as \textit{confluent coarsening}, wherein droplets move advectively along gradients in the height of the film towards low-laying regions of the film.}
    \label{fig:4_Schematic}
\end{figure}

Our findings show that during the early stages of phase separation, hydrodynamic and diffusive transport modes decouple. During these early stages the phase separation kinetics is dictated by diffusive and evaporative mass transport, in agreement with the findings of Schaefer and collaborators who neglect hydrodynamics altogether~\cite{Schaefer2015StructuringEvaporation,Schaefer2016StructuringEvaporation}. Demixing typically occurs under off-critical conditions and the emergent morphology just after demixing resembles a dispersion of solute-rich droplets in a solvent-rich majority phase. This is a consequence of solvent evaporation gradually destabilizing the solution starting from a (very) low solute concentration. The morphology remains off-critical during the late stages of the demixing process where several coarsening regimes present themselves. These we illustrate schematically in Fig.~\ref{fig:4_Schematic}, showing a side view of the film with the solution-gas interface in blue, the solute-rich phase in gray and the interfaces separating the solute-rich and solute-poor domains in orange. Each coarsening mechanism shown is associated with one or more of the mass transport processes present in our model description. If hydrodynamic transport and solvent evaporation are slow relative to diffusive transport, we find a diffusive (Ostwald-type) coarsening mode as depicted in Fig.~\ref{fig:4_Schematic}A. If evaporation is rapid, an evaporative coarsening mode emerges depicted in Fig.~\ref{fig:4_Schematic}B. The decreasing height of the film results in lateral redistribution of the material in the solute-rich domains. This, in combination with the effect of hydrodynamic interactions between the solute-rich domains that promote coalescence, produces an evaporation-induced coalescence pathway. One of the (attractive) hydrodynamic interactions that promotes coalescence of nearby droplets is the ``compositional'' Marangoni effect that is illustrated in Fig.~\ref{fig:4_Schematic}C. This effect is a result of gradients in the liquid-liquid interfacial tension that itself appears to originate from diffusive mass fluxes between droplets due to Ostwald-type transport~\cite{Shimizu2015AMixtures}. Surprisingly, we also find a, as far as we are aware, novel hydrodynamic coarsening mechanism for off-critical mixtures that we refer to as \textit{confluent coarsening} and is illustrated in Fig.~\ref{fig:4_Schematic}D. The physical origins of this coarsening mode lie in the interaction between the liquid-liquid phase boundaries and the solution--gas surface. This interaction induces a directional motion of the solute-rich domains aligned with gradients in the height of the film where the low-laying regions act as focal points for the droplets to accumulate, again promoting domain coalescence.

The remainder of this Chapter is structured as follows. In Section~\ref{sec:theory}, we present our model and in Section~\ref{sec:results} the results of our numerical calculations. The early-time behavior that emerges from  our model we discuss in detail in Section~\ref{sec:early}, showing that for an initially homogeneous film the compositional evolution is in essence not affected by hydrodynamic transport. We subsequently return to the late-stage coarsening dynamics in Section~\ref{sec:coarsening}, unveiling a novel hydrodynamic coarsening pathway originating from the coupling of the hydrodynamics due to the (curved) solution--gas surface and those due to bulk demixing. Finally, we discuss and conclude our work in Sec.~\ref{sec:discussion_conclusion}.

\section{Theory}\label{sec:theory}
We consider the isothermal, evaporation-driven spinodal decomposition of an incompressible binary solution comprising of a solute and a volatile solvent. We focus on the deposition conditions present during the meniscus-guided deposition of a fluid at high substrate velocities deep in the so-called Landau-Levich regime, complementing our earlier work on phase separation in the evaporative regime~\cite{deBruijn2024PeriodicDeposition}. The film dries far removed from the capillary zone near the fluid inlet where the solution-gas interface is (nearly) parallel with the substrate and the fluid stationary with respect to the (moving) substrate. We therefore use the equivalent situation of a (stationary) solution confined between a stationary substrate and an initially flat solution-gas interface. 

In our model we neglect inertial effects, which is justified as Reynolds numbers in thin films are typically much smaller than unity. Both the solute and solvent are assumed to be neutral with respect to both the substrate and the solution-gas interface, implying that (i) we ignore preferential interactions with either surface, and (ii) the solution-gas surface tension is independent of the composition. Hence, we also neglect Marangoni effects associated with gradients in the surface tension of the free surface. Additionally, we suppress stratification in the film itself. This is reasonable if (i) the height of the film $h \equiv h(x,y,t)$, defined as the distance between the substrate and the fluid-gas interface, is smaller than the characteristic size of the demixed structures and (ii) evaporation is sufficiently slow, that is, if $h k/D_\mathrm{coop} \ll 1$, where $k$ is the velocity with which the height of the fluid-gas interface decreases due to solvent evaporation and $D_\mathrm{coop}$ the cooperative or mutual diffusion coefficient of the solute~\cite{Schaefer2017DynamicSolutions,Larsson2022QuantitativeFilms,Naraigh2010NonlinearFilms}.

If the slope of the height of the film remains small, $|\nabla h| \ll 1$, hydrodynamic mass transport can be described invoking a height-averaged approach also known as the lubrication approximation~\cite{Oron1997Long-scaleFilms}. In this approximation, the Stokes equations reduce to a (generalized) diffusion- or Cahn-Hilliard-type equation for the height of the film, simplifying the description considerably~\cite{Oron1997Long-scaleFilms}. Within this framework, the height-averaged advective and diffusive mass currents can be expressed via gradients in the pressure $p$ and the density in the exchange chemical potential $\Delta \mu$ (in units of energy per volume)~\cite{Thiele2016GradientSurfactant}. We obtain these driving forces as the functional derivatives of a free energy functional $\mathcal{F}[h,\psi]$ with respect to the height $h$ and the ``solute height'' $\psi = h \phi$, where $\phi \equiv \phi(x,y,t)$ is the solute volume fraction~\cite{Mitlin1993DewettingDecomposition,Naraigh2007DynamicalFilms,Naraigh2010NonlinearFilms,Thiele2016GradientSurfactant}. Note that the correct expression for the exchange chemical potential density $\Delta \mu$ cannot be obtained via the functional derivative with respect to the solute volume fraction $\phi$ because $\phi$ implicitly depends on the height as $\phi \propto h^{-1}$. Hence, $\phi$ and $h$ cannot be varied independently in the variational sense. Defining $\Delta \mu$ via the functional derivative with respect to the solute height $\psi$ effectively addresses this issue~\cite{Thiele2016GradientSurfactant}. Parenthetically, since $\psi$ is a height, we may also interpret the exchange chemical potential density $\Delta \mu$ as the partial pressure of the solute.

The free energy functional $\mathcal{F}[h,\psi]$ for our model system can be expressed as the sum of two contributions: one associated with the free solution-gas interface, which is also known as the effective interfacial Hamiltonian~\cite{deGennes2004CapillarityPhenomena,Thiele2016GradientSurfactant}, and the other describing the bulk solution. It is given by
\begin{equation}\label{eq:theory:free_energy}
    \begin{split}
   \mathcal{F}[h,\psi] =  \int \dint{\mathbf{r}} \bigg[\frac{\gamma}{2}|\nabla h|^2 + g(h) + \\ 
    \Delta f h \left(f(\phi) + \frac{\kappa}{2}|\nabla \phi|^2\right)\bigg],
   \end{split}
\end{equation}
where we integrate over the substrate area in the $x$--$y$ plane and $\nabla = (\partial_x, \partial_y)^{T}$ is the two-dimensional lateral gradient operator. We express Eq.~\eqref{eq:theory:free_energy} in terms of the solute volume fraction $\phi = \psi/h$ instead of the solute height $\psi$ for notational convenience.
The first two terms in Eq.~\eqref{eq:theory:free_energy} describe the free surface of the liquid solution. The first term represents the work required to deform the interface, where $\gamma$ is the surface tension, assumed to be independent of temperature and solute concentration~\cite{deGennes2004CapillarityPhenomena,Thiele2016GradientSurfactant}. Hence, we suppress the thermal and so-called solutal Marangoni effects associated with the free surface. The compositional Marangoni effect that finds its origin in gradients in the liquid-liquid interfacial tension, illustrated in Fig.~\ref{fig:4_Schematic}C, remains active. The second term in Eq.~\eqref{eq:theory:free_energy} accounts for the disjoining pressure, arising from the differences in the Van der Waals interactions between the solution and the substrate on the one hand and the gas phase and the substrate on the other. We define it as
\begin{equation}\label{eq:disj}
    g(h) = - \frac{A_\mathrm{H}}{12 \pi h^2},
\end{equation}
with $A_\mathrm{H}$ Hamaker's constant. For the sake of simplicity, we ignore slope and curvature corrections to the disjoining pressure~\cite{Dai2008DisjoiningFilms}. Contributions to Eq.~\eqref{eq:disj} that would allow for the formation of a (stable) precursor film are not included. This also means that our model cannot correctly account for dewetting, which we deem to be outside the scope of this work.

The remaining terms in Eq.~\eqref{eq:theory:free_energy} account for the properties of the bulk mixture, assuming that the composition remains vertically uniform. We introduce $\Delta f= k_\mathrm{B} T/b^3$ as the unit of (free) energy density, where $k_\mathrm{B}$ is Boltzmann's constant, $T$ the absolute temperature and $b^3$ a microscopic volume that enters our model for dimensional consistency. For the (dimensionless) bulk free energy density $f(\phi)$, we adopt the Flory-Huggins model
\begin{equation}\label{eq:theory:FloryHuggins}
    f(\phi) = \phi \ln \phi + (1- \phi) \ln (1-\phi) + \phi(1-\phi) \chi,
\end{equation}
where $\chi$ is the well-known Flory interaction parameter~\cite{Flory1942ThermodynamicsSolutions,Huggins1942SomeCompounds.}. The solution phase separates if the solvent quality is sufficiently low, that is, for $\chi > 2$, and if the volume fraction of solute is somewhere between the low and high-concentration branches of the binodal~\cite{Flory1942ThermodynamicsSolutions}.

The square-gradient contribution $\kappa/2|\nabla \phi|^2$ in Eq.~\eqref{eq:theory:free_energy} penalizes the formation of interfaces between the phases in solution. The ``interfacial stiffness'' $\kappa$ we take as a free and constant parameter for simplicity albeit that in reality it may be a function of the volume fraction $\phi$, the Flory interaction parameter $\chi$ and the molecular weight of the solute for polymeric solutions~\cite{deGennes1980DynamicsBlends,Debye1959AngularMixtures}. Interfaces between the solute and solvent-rich domains carry an interfacial tension $\sigma \propto \Delta f\sqrt{\kappa} (\Delta \phi)^{3/2} \left(\chi - \chi_\mathrm{s}(\phi)\right)^{2/3}$, where $\Delta \phi$ is the difference between the equilibrium concentrations in the solute-rich and solvent-rich phases, and $\chi_\mathrm{s} = \frac{1}{2}\langle\phi\rangle(1-\langle\phi\rangle)$ the Flory interaction parameter at the spinodal for a given (mean) solute concentration $\langle\phi\rangle$~\cite{Cahn1958FreeEnergy,Cahn1959FreeFluid,Anderson1998DIFFUSE-INTERFACEMECHANICS,Konig2021Two-dimensionalMixtures}. Strictly speaking, this expression for the interfacial tension is valid only near the critical point, although it seems to remain very accurate far from it~\cite{Konig2021Two-dimensionalMixtures}. In volatile mixtures the mean solute concentration increases with time, rendering the interfacial tension between solute-rich and solute-poor domains in the film, $\sigma$, time dependent as well. Notably, it vanishes when the mean concentration equals the concentration of either branch of the spinodal.

Following the standard approach, we next utilize Onsager's reciprocal relations to relate the time evolution equations for the order parameters $h$ and $\psi$ to the diffusive and hydrodynamic mass currents~\cite{Onsager1931ReciprocalI.,Onsager1931ReciprocalII.}. This yields
\begin{equation}\label{eq:theory:evolution_height}
    \dpart{h}{t} = \nabla\cdot\frac{h^3}{3 \eta} \left(\nabla p + \phi \nabla\Delta\mu\right) + f_\mathrm{evap}(\phi),
\end{equation}
for the height of the film and
\begin{equation}\label{eq:theory:evolution_partial_height}
    \dpart{\psi}{t} = \nabla\cdot \frac{h^2\psi}{3 \eta} \left( \nabla p + \phi \nabla \Delta\mu\right) + \nabla \cdot h M(\phi) \nabla \Delta\mu + \zeta,
\end{equation}
for the solute height. Here, we introduce the known expressions for Onsager's mobility coefficients with $\eta$ the viscosity of the solution that we assume to be constant for simplicity~\cite{Mitlin1993DewettingDecomposition,Xu2015AMixtures,Thiele2016GradientSurfactant}. The diffusive mobility for an incompressible binary mixture reads~\cite{Doi2011OnsagersMatter}
\begin{equation}\label{eq:theory:mobility}
    M = \Delta f^{-1} D \phi (1- \phi),
\end{equation}
which is also known as the ``double degenerate'' mobility~\cite{Dai2016ComputationalMobility}, with $D$ the tracer (self) diffusivity that we assume to be constant. The evaporation flux $f_\mathrm{evap}(\phi)$ and thermal noise $\zeta$ we return to below. The pressure is given by $p = \delta \mathcal{F}/\delta h = -\gamma \nabla^2 h + \partial_h g(h) + p_\mathrm{b}$ with $p_\mathrm{b} = \Delta f (f(\phi) + \kappa/2 |\nabla \phi |^2) - \phi \Delta\mu$ the osmotic pressure of the solution. The exchange chemical potential density is defined as $\Delta \mu = \delta \mathcal{F}/\delta \psi = \Delta f \partial_\phi f(\phi) - \Delta f h^{-1} \kappa \nabla \cdot h \nabla \phi$. In principle, we can now also derive from Eqs.~\eqref{eq:theory:evolution_height} and \eqref{eq:theory:evolution_partial_height} an evolution equation for the solute volume fraction $\phi$. We opt to not do so as the equation for the solute height Eq.~\eqref{eq:theory:evolution_partial_height} is simpler to implement numerically.

As is usual in the lubrication theory of thin films, we interpret 
\begin{equation}\label{eq:theory:velocityfield}
\begin{split}
    \mathbf{u} &\equiv -\frac{h^2}{3 \eta} \left( \nabla p + \phi \nabla\Delta\mu\right) \\ 
    &= -\frac{h^2}{3 \eta} \left[ \nabla (p-p_\mathrm{b}) + \Delta f \kappa (\nabla| \nabla \phi |^2 + (h^{-1} \nabla h \cdot\nabla \phi)\nabla \phi)\right]
\end{split}
\end{equation}
 as the height-averaged fluid velocity~\cite{Thiele2016GradientSurfactant,Oron1997Long-scaleFilms}. For the second equality sign we use $\nabla p_\mathrm{b} = - \phi \nabla\Delta\mu +  \Delta f \kappa [\nabla| \nabla \phi |^2 + (h^{-1} \nabla h \cdot\nabla \phi)\nabla \phi]$, which follows by taking the gradient of the osmotic pressure $p_\mathrm{b}$ where we make use of the expression for the exchange chemical potential $\Delta \mu$~\cite{Thiele2016GradientSurfactant}. We are thus led to conclude that the fluid velocity must be independent of the osmotic pressure $p_\mathrm{b}$ but that it does depend on the presence of interfaces between the solute-rich and solvent-rich phases. As we shall show at a later stage of this work, the final contribution $\Delta f \kappa (h^{-1} \nabla h \cdot\nabla \phi)\nabla \phi$ to Eq.~\eqref{eq:theory:velocityfield} can result in directional motion of droplets if the height of the film has a gradient. 
 
Next, for the solvent evaporation flux $f_\mathrm{evap}(\phi)$, we use the simple ansatz of a linear relation between the solvent concentration at the solution-gas surface and the evaporation flux 
\begin{equation}\label{eq:theory:evaporation}
    f_\mathrm{evap}(\phi) = - k(1-\phi).
\end{equation}
Here, $k$ is a phenomenological mass-transfer coefficient that depends on the partial pressure of the solvent, the solvent quality, and so on~\cite{Bornside1989SpinModel}. 

Finally, as usual the thermal noise $\zeta$ is delta-correlated with zero mean $\langle \zeta(x,y,t)\rangle = 0$ and covariance $\langle \zeta(x,y,t) \zeta(x',y',t') \rangle = -2 \kBT \omega^2 \nabla\cdot h M(\psi,h) \nabla\delta(x-x')\delta(y-y')\delta(t-t')$. Here, $\omega\leq 1$ is an ad hoc scaling parameter~\cite{Cook1970BrownianDecomposition,Ronsin2022PhaseFieldFilms} that has no physical origin but dampens the intensity of the thermal fluctuations. This allows us to take larger steps in our time integrator. For $\omega \neq 1$ the  \textit{magnitude} (in some sense) of the noise violates the fluctuation-dissipation theorem, yet we find justification for setting $\omega < 1$ in the observation that it affects our results only quantitatively not qualitatively, and that thermal fluctuations can generally anyway be neglected in the late times of coarsening~\cite{Puri1988EffectDecomposition,Konig2021Two-dimensionalMixtures}. We do not account for any thermal fluctuations in the height of the film in Eq.~\eqref{eq:theory:evolution_height}~\cite{Clarke2005TowardMixtures,Davidovitch2005SpreadingFluctuations,Grun2006Thin-FilmNoise} nor for any cross-correlated thermal fluctuations, and discuss the consequences of these approximations at a later stage in this Chapter.

To make our model description as generic as possible, we nondimensionalize our model using the initial height of the film $h_0$ as the characteristic scale for the height and also to nondimensionalize the lateral lengths. For the velocity scale we take $u = \Delta f \sqrt{\kappa}/\eta \propto \sigma/\eta$, because $\Delta f \sqrt{\kappa}$ is a measure for the interfacial tension between solute-rich and solute-poor domains, $\sigma$~\cite{Cahn1958FreeEnergy,Cahn1959FreeFluid,Konig2021Two-dimensionalMixtures}; see also our discussion earlier in this section. The pressure scale we define as $p_0 =\eta u/h_0$ and the diffusive time scale as $t_0 = h_0^2/D$. The relevant dimensionless groups are (i) the Capillary number $\Ca \equiv u \eta/\gamma \propto \sigma/\gamma$, which also acts as a measure for the ratio of the interfacial tension between the solute-rich and solute-poor regions and that of the fluid-gas interface, (ii) what we call the disjoining number $\G = A_\mathrm{H}/ 6 \pi h_0^2  \eta u$, which measures the strength of the disjoining forces relative to the capillary forces of the solute-solvent interfaces, (iii) the Peclet number $\Pe = u h_0/D = \Delta f \sqrt{\kappa} h_0/ \eta D$, (iv) the Biot number $\mathrm{Bi} =  k h_0/D$, which measures the strength of evaporation relative to diffusion, and (v) the Cahn number $\Cn = \kappa/h_0^{2}$. For the remainder of this work we treat the Peclet number, the Biot number and the Capillary number as freely adjustable parameters. We insert the dimensionless variables and operators $h = h/h_0$, $\psi = \psi/h_0$, $p = p/p_0$, $\Delta\mu = \Delta\mu/\Delta f$, $\nabla = h_0 \nabla$, $t = t/t_0$, $M = M \Delta f/D$ and $\zeta = \zeta \Delta f/D h_0$ in the governing equations Eqs.~\eqref{eq:theory:evolution_height}-\eqref{eq:theory:evaporation}, producing the following dimensionless equations for the film height
\begin{equation}\label{eq:theory:dimless:height}
    \dpart{h}{t} = \Pe \nabla\cdot  \frac{h^3}{3} \left( \nabla p + \frac{1}{\sqrt{\Cn}}\phi\nabla\Delta\mu\right) - \mathrm{Bi}(1-\phi),
\end{equation}
with the pressure 
\begin{equation}\label{eq:theory:dimless:pressure}
    p = -\frac{1}{\Ca} \nabla^2 h + G/h^{3} + \frac{1}{\sqrt{\Cn}} p_\mathrm{b},
\end{equation}
and for the solute height
\begin{equation}\label{eq:theory:dimless:partial_height}
\dpart{\psi}{t} = \Pe \nabla\cdot \frac{h^2\psi}{3} \left( \nabla p + \frac{1}{\sqrt{\Cn}}\phi\nabla\Delta\mu\right) +\nabla \cdot h M \nabla \Delta \mu + \zeta,
\end{equation}
with the exchange chemical potential
\begin{equation}\label{eq:theory:dimless:chem_pot}
    \Delta \mu = \partial_\phi f(\phi) - \frac{\Cn}{h} \nabla \cdot h \nabla \phi.
\end{equation}

We solve our model equations numerically using for the physical parameters and dimensionless numbers the values listed in Table~\ref{tab:theory:parameter_waardes}. For the concentration gradient stiffness $\kappa$ we use that for organic semiconductors $\Delta f \kappa$ is generally estimated to be on the order of $10^{-10} - 10^{-12}$ J/m~\cite{Saylor2007Diffuse-interfaceSystems,Wodo2014HowBlends,Clarke2005TowardMixtures}. Using an order-of-magnitude estimate for the microscopic volume $b^3 = 10^{-28}$ m$^{3}$ and the thermal energy $\kBT = 10^{-21}$ J we find $\kappa \approx \mathcal{O}(10^{-1} - 10^{1})$ nm$^{2}$. Our choice for $\kappa = 25$ nm$^{2}$ is to ensure a sufficient number of grid points in the phase boundaries, while still within the range of reasonable values. We discretize the gradient operators using second-order central finite differences and the contribution of the thermal noise using the method of Schaefer \textit{et al.}~\cite{Schaefer2016StructuringEvaporation}. Time is integrated making use of a semi-implicit Euler time integrator, integrating the thermal noise $\zeta$ explicitly and all other terms implicitly. Our method conserves the solute mass up to negligible numerical errors of the order of $< 10^{-7} \%$ between the initial and final solute mass. Adaptive time steps are employed following the approach outlined by Wodo and Ganapathysubramanian~\cite{Wodo2011ComputationallyProblem}. We invoke periodic boundary conditions and initialize our calculations with a homogeneous and flat thin film, setting the initial volume fraction equal to the low concentration branch of the spinodal. This is reasonable as the metastable region is typically traversed in experimental situations due to fast evaporation. The model is implemented in parallel using the PETSc library~\cite{Abhyankar2018PETSc/TS:Library,*Balay2024PETSc/TAOManual,*Balay2024PETScPage}.

In the next sections, we first present and discuss the phenomenology of our numerical calculations and discuss how the effective evaporation rate depends on the Peclet number. Subsequently, we discuss in detail the early stages of demixing in our model, demonstrating that hydrodynamic and diffusive transport modes decouple. Consequently, during the early stages of demixing, diffusive and evaporative transport dominate. Finally, we examine the late stage coarsening of the mixture across a range of values of the Peclet, Biot and Capillary numbers. As already advertised, we identify a novel coarsening mechanism driven by the interaction of the fluid-fluid interfaces with the fluid-gas interface.

\begin{table}[tb]
    \centering
    \caption{A list of parameter values used in this Chapter.}
    \label{tab:theory:parameter_waardes}
    \begin{tabular}{| c | c | c |}
        \hline
        Parameter & units & value \\
        \hline
        $h_0$ & [nm] & $30$ \\
        $\phi_0$ & [ - ] & $0.1464467\dots$ \\
        $\gamma$ & [mN/m] & $25$ \\
        $D$ & [m$^2$/s] & $10^{-10}$ \\
        $k$ & m/s & $\mathcal{O}(10^{-6} - 10^{-3})$\\
        $A_\mathrm{H}$ & J & $10^{-19}$ \\
        $\kappa$ & [nm$^{2}$] & 25 \\
        $\chi$ & [ - ] & 4 \\ \hline
        $\omega$ & [ - ] & $10^{-2}$ \\
        $\Ca$ & [ - ] & $\mathcal{O}(10^{-3} - 10^{-1})$ \\
        $\Pe$ & [ - ] & $\mathcal{O}(10^{-2} - 10^{2})$ \\
        G  & [ - ] & $7 \times 10^{-2}$ \\
        $\Cn$ & [ - ] & $2.66 \times 10^{-2}$ \\
        $\Bi$ & [ - ] & $\mathcal{O}(10^{-3} -  10^{-1})$ \\ \hline
    \end{tabular}

\end{table}

\section{Model calculations}\label{sec:results}
In this section we present and discuss our numerical results for demixing taking place in a binary fluid film containing a solute and a volatile solvent. As already alluded to, our results apply to (i) stationary films and (ii) films deposited on a rapidly moving substrate, \textit{i.e.}, deep in the Landau-Levich regime during meniscus-guided deposition, wherein the mixture is, for all intends and purposes, stationary with respect to the moving substrate. We solve Eqs.~\eqref{eq:theory:dimless:height}--\eqref{eq:theory:dimless:chem_pot} for a host of parameter values listed in Table~\ref{tab:theory:parameter_waardes}. Because the fluid-gas surface can freely respond to the formation of domains, and because of the difference in evaporation rates in the solute and solvent-rich phases, the effective evaporation rate depends not only on the Biot number but also on the other parameters of our model. Hence, we also investigate how the other parameters affect the evaporation and demixing kinetics.

\begin{figure}
    \centering
    \includegraphics[width=\columnwidth]{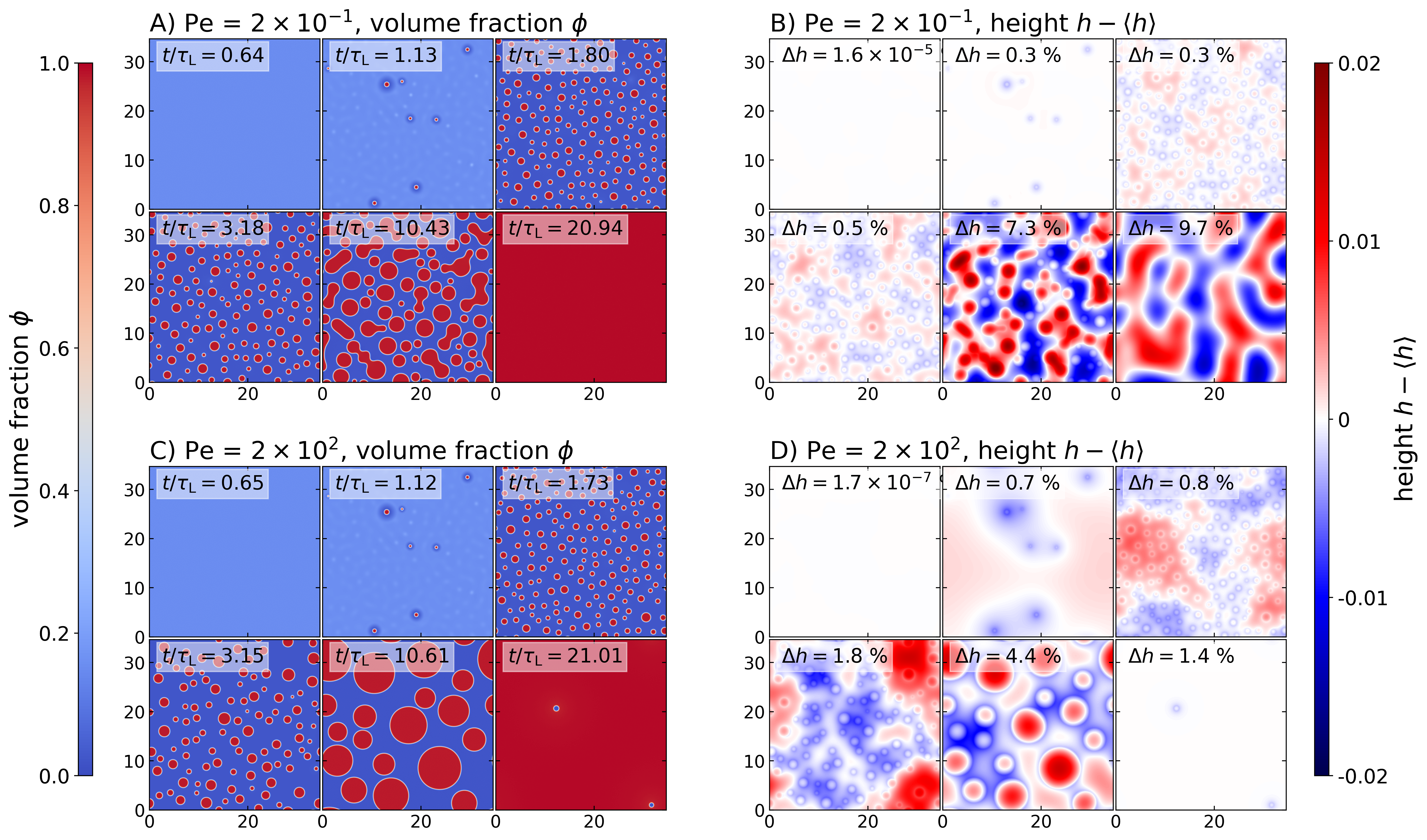}
    \caption{Snapshots of the volume fraction (A and C) and the height of the film (B and D) for two different values of the Peclet number at different moments in time $t/\tau_\mathrm{L}$. The time is scaled to the spinodal lag time $\tau_\mathrm{L}$, see the main text. Panels A and B are for $\Pe = 2 \times 10^{-1}$, Panels C and D for $\Pe = 2 \times 10^{2}$. The left color bar indicates the volume fractions for panels A and C, the right color bar is $h - \langle h \rangle$ for panels B and D. $\Delta h = 100 \times \text{max}(| h - \langle h \rangle|/\langle h \rangle)$ is a measure for the relative roughness of the film surface. Other model parameters are $\Cn = 0.0267$, $\Ca = 2 \times 10^{-2}$, $\Bi = 3\times 10^{-3}$, $\chi = 4$, and $G = 2.3 \times 10^{-4}$.}
    \label{fig:snapshots}
\end{figure}

Fig.~\ref{fig:snapshots} shows representative snapshots of the local volume fraction and height of the film relative to the mean height in a square domain using periodic boundary conditions, for different times and with a Peclet number $\Pe = 2 \times 10^{-1}$ in Fig.~\ref{fig:snapshots}A and Fig.~\ref{fig:snapshots}B and with $\Pe = 2 \times 10^{2}$ in Fig.~\ref{fig:snapshots}C and Fig.~\ref{fig:snapshots}D. We set $\Bi = 3 \times 10^{-3}$, $\Ca = 2\times 10^{-2}$, and the other parameters as listed in Table~\ref{tab:theory:parameter_waardes}. The indicated times $t$ are scaled to the spinodal lag time $\tau_\mathrm{L}$, also known as the spinodal amplification time~\cite{Binder1983CollectiveMixtures}. This is the waiting time between crossing the spinodal at $t=0$ and the moment in time that the mixture phase separates appreciably~\cite{Schaefer2015StructuringEvaporation,Schaefer2016StructuringEvaporation}. Note that we use the same \textit{seed} for our random number generator for both values of the Peclet number shown for the sake of comparison. This results in (nearly) identical integrated thermal noise for both values Peclet numbers, although the one-to-one correspondence is eventually destroyed due to the adaptivity of our numerical time stepper.

After crossing the spinodal at $t = 0$, concentration fluctuations remain small for times up to the spinodal lag time $\tau_\mathrm{L}$, as can be seen in Fig.~\ref{fig:snapshots}A and \ref{fig:snapshots}C~\cite{Schaefer2015StructuringEvaporation}. For $t>\tau_\mathrm{L}$, the solution phase separates into solute-rich droplets dispersed in a solvent-rich majority phase, which is typical for off-critical phase separation. In our calculations phase separation tends to occur under off-critical conditions because the mixture is gradually destabilized by solvent evaporation starting from the (very off-critical) low concentration branch of the spinodal. The domains ripen due to solvent evaporation, as well as due to diffusive and hydrodynamic coarsening. The late-stage morphological evolution and coarsening rate appears to be quite sensitive to the Peclet number. We obtain a different morphology and a larger characteristic feature size for the high Peclet number, shown in Fig.~\ref{fig:snapshots}C and D, than for small Peclet numbers, shown in Fig.~\ref{fig:snapshots}A and B. This is actually somewhat surprising considering that hydrodynamic coarsening is generally believed to be of minor importance for off-critical mixtures~\cite{Tanaka1996CoarseningMixtures,Shimizu2015AMixtures}. We return to this in our discussion of Sections~\ref{sec:coarsening} and \ref{sec:flowfields}. Under the action of solvent evaporation the morphology eventually inverts from solute-rich droplets dispersed in a solvent majority phase to a dispersion of solvent-rich droplets  and the solution subsequently redissolves. Redissolution commences upon crossing the high-concentration branch of the binodal and is, of course, a property of the binary solution, whereas for two or more non-volatile components the film typically remains demixed even after the solvent is removed~\cite{Negi2018SimulatingInvestigation,Franeker2015spincoating}. Note that the solution actually already redissolves slightly before crossing the binodal, because the free-energetic cost of the liquid-liquid interface is for the very small solvent-rich droplets not outweighed by the gain in free energy due to phase separation.

Accompanying the morphological evolution of the bulk solution is that of the free surface of the film, as shown in Fig.~\ref{fig:snapshots}B for Peclet number $\Pe = 2\times10^{-1}$ and Fig.~\ref{fig:snapshots}D for $\Pe = 2\times 10^{2}$. The times associated with the sequence of panels in A and B, and of C and D, are the same. Indicated in the panels of B and D are the quantities $\Delta h = 100 \times \text{max}(| h - \langle h \rangle|/\langle h \rangle)$ that measure the roughness of the film relative to the mean height. The surface only deforms after the solution demixes for $t > \tau_\mathrm{L}$, on the one hand due to the forces exerted on it by the fluid-fluid interfaces that develop in the film and on the other due to the spatial gradients in the rate of solvent evaporation at the film surface. The only concentration-dependent (downward) force exerted on the solution-gas interface is the interfacial tension of the liquid-liquid interfaces in the film itself, resulting at the free surface in three-phase liquid-liquid-gas contact lines. The contact angles at the three-phase contact line are always small because the interfacial tension between demixing liquid domains is much smaller than that of the solution-gas interface. This is to be expected~\cite{deGennes2004CapillarityPhenomena}. Different values of the interfacial tension between the solute-rich and solute-poor region and of the fluid-gas interfacial tension, which we achieve by varying the Cahn or Capillary numbers, yield qualitatively comparable results albeit with different solute-solvent-gas contact angles (not shown). 

In the later stages of the drying process before the mixture redissolves, the regions rich in solute decrease less rapidly in height than the regions rich in solvent do as Fig.~\ref{fig:snapshots} show for $t/\tau_L \gtrapprox 10$. This is caused by difference in the evaporation rate between those regions. On the other hand, the Laplace pressure that is a result of the curved liquid-gas interface tends to counteract this, yet cannot immediately compensate for it. We read off from Eqs.~\eqref{eq:theory:dimless:height} and \eqref{eq:theory:dimless:pressure} that the contribution of evaporation to variations in the film thickness relative to that of the material distribution driven by the Laplace pressure of the solution-gas surface -- the first term in Eq.~\eqref{eq:theory:dimless:pressure} -- must scale as $\Bi \times (\Ca/\Pe)$ in terms of the Biot number $\Bi$, Peclet number $\Pe$ and Capillary number $\Ca$. Hence, for constant evaporation rate the magnitude of evaporation-induced surface roughness should increase with decreasing Peclet number and increasing Capillary number. This is in agreement with our results summarized in the snapshots of Figs.~\ref{fig:snapshots}B and \ref{fig:snapshots}D. 
In fact, the bottom right panel of Fig.~\ref{fig:snapshots}B, shows that the surface inhomogeneities may persist even after the solute redissolves and the solution is again of uniform composition. These surface inhomogeneities are in our model not actually frozen in the dry film, because we assume the viscosity to be independent of the composition. Consequently, the liquid-gas interface in the end relaxes to become flat over a wave number $q$ dependent time scale $\tau_\mathrm{h}(q)$ that we estimate to obey the relation 
\begin{equation}
    \tau_\mathrm{h}(q) \sim \frac{1}{\Pe} \left[\frac{G}{h_\mathrm{dry}} q^2 + \frac{h_\mathrm{dry}^3}{\Ca}  q^4\right]^{-1}
\end{equation}
with $h_\mathrm{dry}$ the dry height of the film, $q$ the (dimensionless) wave number and $G$ the disjoining number. This estimate can be obtained from Eq.~\eqref{eq:theory:dimless:height} by calculating the linear response of the height of the film to a periodic perturbation of wave number $q$ around the dry (solvent-free) height of the film $h_\mathrm{dry}$. Apparently, the lifetime of the free-surface roughness increases with decreasing Peclet number, a prediction that is in agreement with our findings of Fig.~\ref{fig:snapshots}.

\begin{figure}
    \centering
    \includegraphics[width=\columnwidth]{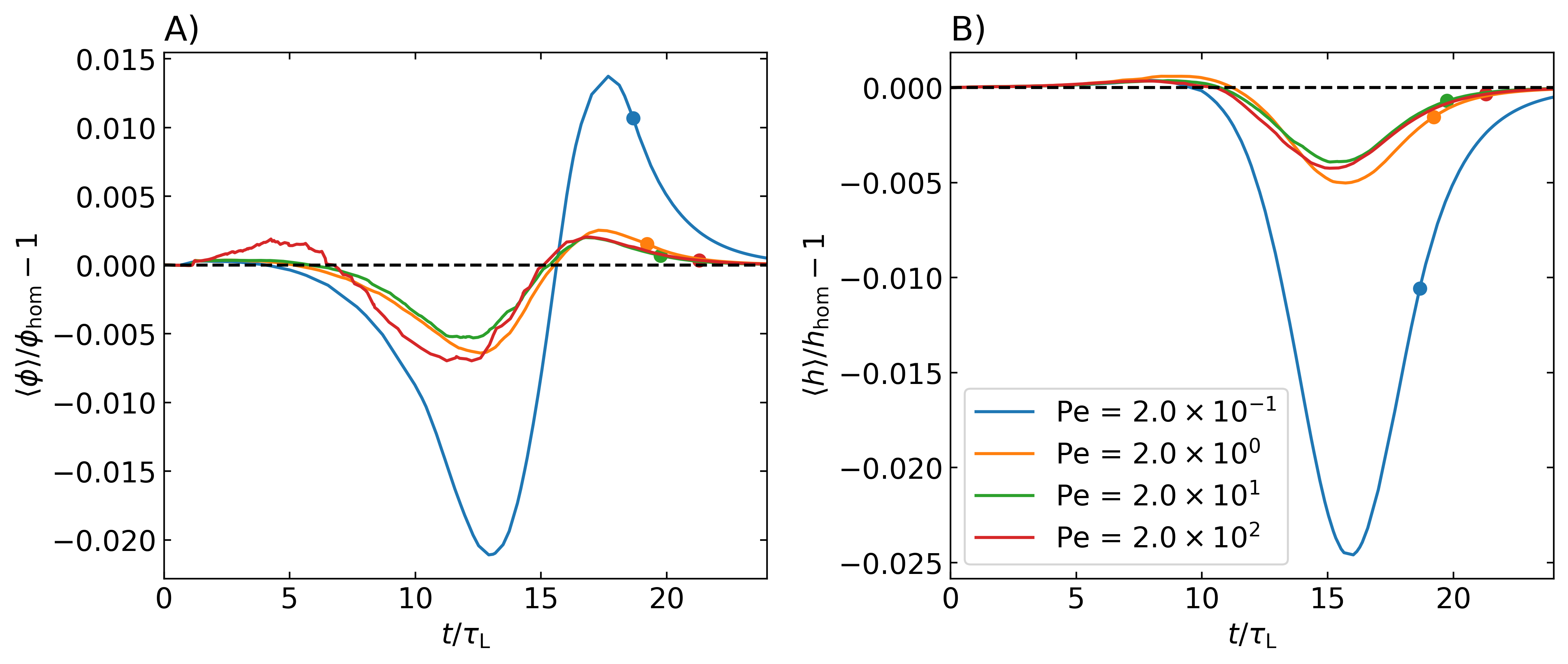}
    \caption{The drying of a demixing solution compared to the drying of a non-demixing, homogeneous film for four values of the Peclet number indicated in the legend of panel B). In Panel A we show the deviation of the area-averaged volume fraction $\langle \phi \rangle$ from the volume fraction $\phi_\mathrm{hom}$ of a drying film that remains homogeneous, as a function of the scaled time $t/\tau_\mathrm{L}$ with $t=\tau_\mathrm{L}$ the moment in time that the film starts to demix. The dots on the curves indicate the time at which the binary mixture redissolves. The horizontal dashed line is a guide for the eye. B) The deviation of the area-averaged height $\langle h \rangle$ of the film relative to that of a film that remains homogeneous $h_\mathrm{hom}$, as a function of the scaled time. The values for the model parameters are $\Cn = 0.0267$, $\Ca = 2 \times 10^{-2}$, $\Bi = 3\times 10^{-3}$, $\chi = 4$, $\phi(t=0) = \psi(t=0) = 0.1464$ and $G = 7 \times 10^{-2}$. The curves in A) and B) with the largest variation of the volume fraction and height are those with the lowest Peclet number, $Pe=0.2$}
    \label{fig:dryingstatistics}
\end{figure}

Upon comparing the final snapshots in Fig.~\ref{fig:snapshots}A and Fig.~\ref{fig:snapshots}C, it transpires that the solution redissolves earlier for $\Pe = 2\times 10^{-1}$ than for $\Pe = 2\times 10^{2}$. Hence, the \textit{effective} drying or evaporation rate depends not only on the Biot number but also on the other dimensionless numbers entering our model. How precisely, should be somewhat sensitive to the evaporation model used. For our particular evaporation model, expressed in Eq.~\eqref{eq:theory:evaporation}, the reason that the drying rate depends on the other dimensionless numbers entering our model is that the overall drying rate of the phase-separating film is the surface average of that of the solvent- and solute-rich domains. The former contribute more per unit area on account of the lower volume fraction of solute. The fraction of the substrate that is covered by the solvent-rich and poor phases also depends on the difference in their height. The film turns out to be thicker in the solute-rich regions than in the solvent-rich regions, and this difference increases with decreasing Peclet number and with increasing Capillary number. Consequently, the fraction of the substrate covered by the solvent-rich phase and therefore the effective evaporation rate must also increase with decreasing Peclet number and increasing Capillary number. 

While this explains why the drying time not only depends on the Biot number but also on the other dimensionless groups of our model, this does not explain why we find that the actual time-resolved drying kinetics exhibits intervals where the evaporation either speeds up or slows down. To quantify this, we compare the time-dependent area-averaged concentration $\langle\phi\rangle(t)$, and the area-averaged height $\langle h \rangle(t)$ to those in a film of homogeneous composition and same initial volume fraction of the solute wherein solvent evaporates according to Eq.~\eqref{eq:theory:evaporation}. In Fig.~\ref{fig:dryingstatistics}A we compare the area-averaged volume fraction $\langle\phi\rangle(t)$ with the concentration $\phi_\mathrm{hom}(t)$ in a film that remains homogeneous, \textit{i.e.}, with Flory interaction parameter $\chi = 0$, for four different values of the Peclet number between $\Pe = 2\times 10^{-1}$ and $\Pe = 2\times10^{2}$ for a Biot number of $\Bi = 3 \times 10^{-3}$ and a capillary number of $\Ca = 5 \times 10^{-2}$. The (color-matched) circles on the curves indicate the moment in time that the solution redissolves, which indeed shifts to earlier times for decreasing Peclet number, again, this is a consequence of the effect of the Peclet number on the difference in the height of the regions rich in either solute or solvent. Positive values indicate that solvent evaporation is slower in the demixed film than in a corresponding homogeneous film and negative values that it is faster. For $t < \tau_\mathrm{L}$ both films are homogeneous and therefore dry at an identical rate. For $t > \tau_\mathrm{L}$, we find that the evaporation rate first increases a little with time to decrease significantly and subsequently increase and finally to decrease again. So, there seem to be two maxima separated by a minimum. The under- and overshoot for the small Peclet number equal to $\Pe = 2\times 10^{-1}$ decreases much more rapidly than the other curves, which we explain below. Note also that the primary maximum increases with increasing Peclet number, which is a consequence of the variations in the height being somewhat larger at high than at low Peclet number during the early stages of demixing, as can be read off Fig.~\ref{fig:snapshots}B and Fig.~\ref{fig:snapshots}D.

In Fig.~\ref{fig:dryingstatistics}B, we compare the mean height of the film to the height of a corresponding homogeneous film. In agreement with our observations from Fig.~\ref{fig:dryingstatistics}A, we conclude that the height of the film is initially larger than that in the homogeneous film, subsequently decreases and remains below that of a homogeneous film. For later times the height is smaller than that of a homogeneous film, consistent with our argument that the effective evaporation rate is faster for a demixed film with a deformed surface than for a homogeneous film with a flat solution-gas surface. The curve for $\Pe = 2\times10^{-1}$ decreases much more rapidly than for the other values of the Peclet number. The reason is that for small values of the ratio $\Pe/\Ca$ the effect of solvent evaporation is stronger than that of hydrodynamic material redistribution. This results in larger differences in the height of the solute-rich and solute-poor domains and consequently also in larger differences in the mean volume fraction in agreement with our findings presented in Fig.~\ref{fig:dryingstatistics}A. For even lower values of the ratio $\Pe/\Ca$, we in fact find an evaporation-induced dewetting transition. We deem this outside the scope of the present Chapter and therefore do not study this in detail.

Having discussed the phenomenology of the demixing and the drying of the film, we next investigate, separately, the early and late stages of demixing. First, in the following section, we study the early stages of demixing up to the moment in time that the solution phase separates, at $t = \tau_\mathrm{L}$. Here, we interestingly find that solvent evaporation has a strong impact on the early stage temporal evolution of the volume fractions. Following this, we discuss the late-stage coarsening and put forward an explanation for the differences in structural evolution observed in Fig.~\ref{fig:snapshots} for low and high Peclet numbers, and identify a, as far as we are aware, novel coarsening mechanism. This what we earlier in this work refer to as \textit{confluent coarsening} is the results of a coupling between the bulk and surface hydrodynamic transport modes.

\section{Early stage behavior}\label{sec:early}
Let us now focus on the growth of density fluctuations in the pre-demixing stage, and investigate the linear response of the height and volume fraction fields to a thermal excitation. For non-volatile mixtures this pre-demixing stage has already been investigated by Clarke~\cite{Clarke2005TowardMixtures} and Naraigh \textit{et al}~\cite{Naraigh2010NonlinearFilms} showing that the height and volume fractions evolve independently if the disjoining pressure and liquid-vapor surface tension are independent of solute concentration. Moreover, the temporal evolution of the volume fraction was found to be diffusive and unaffected by hydrodynamic transport, which is consistent with predictions for bulk models where hydrodynamic transport becomes important only after the liquid-liquid phase boundary become sufficiently sharp~\cite{Chen1998HydrodynamicBlends,Tanaka1996CoarseningMixtures}. As we show next, volatile mixtures differ considerably from non-volatile mixtures, because the height and volume fraction fields couple via solvent evaporation. Nevertheless, we argue that because we neglect thermal fluctuations in the height field, this coupling turns out to be weak and can be disregarded in our numerical calculations.

In order to characterise the pre-demixing stage, we seek to extract the delay in time $\tau_\mathrm{L}$ between crossing the spinodal at $t = 0$ and the moment in time at $t = \tau_\mathrm{L}$ that the solution actually phase separates, as well as the characteristic feature size of the phase separated solution measured in terms of the associated emergent wave number $q_*$. We apply our analysis to the volume fraction field $\phi = \psi/h$ instead of the partial height $\psi$, since the former is the order parameter that best describes the demixing kinetics. To do this, we first recast the equation for the solute height~\eqref{eq:theory:dimless:partial_height} into an evolution equation for the solute volume fraction $\phi$ and subsequently linearize both the height and the volume fraction field around a homogeneous but drying thin film with time-dependent composition $\phi_\mathrm{hom}(t)$ and height $h_\mathrm{hom}(t)$. The set of linearized equations read in Fourier space
\begin{equation}\label{eq:early:linearized}
    \frac{\partial }{\partial t} 
    \begin{pmatrix}
\delta \phi_q \\
\delta h_q
\end{pmatrix}
    = \mathbf{Q} \cdot \begin{pmatrix} 
\delta \phi_q \\
\delta h_q
\end{pmatrix} + 
\mathbf{\zeta}_q
\end{equation}
with $t$ again the dimensionless time, $\delta \phi_q$ and $\delta h_q$ the Fourier transforms of the volume fraction and height fluctuations around $\phi_\mathrm{hom}(t)$ and $h_\mathrm{hom}(t)$ with $q$ the wave number of the fluctuation and $\mathbf{\zeta}_q$ the thermal fluctuations. For simplicity, we assume in our analysis an initial thermal excitation only, and neglect thermal fluctuations for $t>0$.
The matrix of coefficients reads 
\begin{equation}\label{eq:coefficient_matrix}
    \mathbf{Q} = \begin{pmatrix}
        Q_\mathrm{\phi\phi} & Q_\mathrm{\phi h} \\
        Q_\mathrm{h \phi} & Q_\mathrm{h h}
    \end{pmatrix} = 
    \begin{pmatrix}
        R(q,t) + \Bi\left(1-2\phi_\mathrm{hom}\right)h_\mathrm{hom}^{-1}  & -\Bi\phi_\mathrm{hom}\left(1-\phi_\mathrm{hom}\right)h_\mathrm{hom}^{^-2} \\
        \Bi & \frac{1}{3} \Pe \hspace{0.1cm} h_\mathrm{hom}^3 \hspace{0.1cm}q^2 \left(3 \mathrm{G} \hspace{0.1cm} h_\mathrm{hom}^{-4} + \Ca^{-1} q^2\right)
    \end{pmatrix},
\end{equation}
with $R(q,t) = M(t) q^2\left[\phi_\mathrm{hom}^{-1} + (1-\phi_\mathrm{hom})^{-1} - 2 \chi + \Cn q^2\right]$ with $q$ the (dimensionless) wave number and $G$ the disjoining number. Note that the kinetic matrix $Q$ depends on time because the reference state dries too, and is described by a time-dependent volume fraction and film height $\{\phi_\mathrm{hom}(t), h_\mathrm{hom}(t) \}$. Hence, we cannot proceed by the usual linear stability analysis. 

To make headway, let us first note that the first term in the diagonal $Q_\mathrm{\phi\phi}$ component in Eq.~\eqref{eq:coefficient_matrix} accounts for diffusive mass transport via $R(q,t)$ and the second one in $Q_\mathrm{\phi\phi}$ accounts for the effect of the concentration dependence of the solvent evaporation rate. Both off-diagonal terms $Q_\mathrm{\phi h}$ and $Q_\mathrm{h\phi}$ that couple the local volume fraction and the height of the film originate from solvent evaporation only. Hence, in agreement with earlier work that include hydrodynamics, we conclude that hydrodynamic transport modes do not contribute to the initial amplification of the primary unstable spinodal density wave~\cite{Chen1998HydrodynamicBlends,Naraigh2010NonlinearFilms,Clarke2005TowardMixtures,Siggia1979LateMixtures,Tanaka1996CoarseningMixtures,Shimizu2015AMixtures}. The second diagonal contribution $Q_\mathrm{h h}$, describing coupling of fluctuations of the height of the film, accounts for hydrodynamic redistribution of the bulk material. It is interesting to note that the kinetic matrix diagonalizes only for non-volatile mixtures for which $\Bi = 0$, which have been analyzed by Clarke~\cite{Clarke2005TowardMixtures} and Nargaith \textit{et al.}~\cite{Naraigh2010NonlinearFilms}. This, perhaps surprisingly, also suggests that (thermal) fluctuations in the height of the film must have a different effect on the initial phase separation kinetics of volatile and that of non-volatile mixtures. 

We next seek a solution to Eq.~\eqref{eq:early:linearized} to obtain the spinodal lag time $\tau_\mathrm{L}$ and the emergent wave number $q_*$. This is actually not quite straightforward because the kinetic matrix $Q$ is time dependent, as already announced. The standard approach to diagonalize the kinetic matrix $Q$ does not yield the exact solution to Eq.~\eqref{eq:early:linearized}, but only provides a zeroth order contribution to the solution in a so-called Magnus expansion~\cite{Magnus1954OnOperator}. Higher order corrections can then be calculated in terms of the commutator of the kinetic matrix with itself, evaluated at different moments in time. For our model, this commutator is non-zero and therefore higher order terms do not vanish. Instead, we opt to first simplify the problem at hand to reflect our numerical calculations and subsequently solve the remaining equations. First, we reiterate that we neglect in our calculations thermal fluctuations in the height of the film and that the height is initially constant. Fluctuations in the height of the film are therefore excited \textit{indirectly} via the thermal fluctuations in the solute height (or volume fraction). In our numerical calculations, the magnitude of the fluctuations in the height of the film remains many orders of magnitude smaller than the fluctuations in the volume fractions. Hence, we argue that we may neglect the off-diagonal contribution $Q_{\phi h}$. In practise, this means that the volume fraction field evolves independently of the height field, whereas the height field remains affected by and is subservient to the local volume fraction. 

Using these simplifications, we only need to solve the equation for the solute volume fraction to extract the spinodal lag time $\tau_\mathrm{L}$ and the emergent wave number $q_*$. This equation was already analyzed by Schaefer \textit{et al.}~\cite{Schaefer2015StructuringEvaporation}, albeit for a different evaporation model wherein the volume fraction increases linearly with time, in which case the second term in $Q_{\phi\phi}$ in Eq.~\eqref{eq:coefficient_matrix} drops out of the equation. Setting this term to zero is actually also justified in our case because $\Bi \ll 1$ is a necessary condition for our height-averaged model to be valid. Following Schaefer \textit{et al.}~\cite{Schaefer2015StructuringEvaporation}, we introduce a spinodal diffusion time $\tau_\mathrm{d} = \Cn/M(\phi_\mathrm{s})$ and an evaporative destabilization time $\tau_\mathrm{e} = h_0/|f_\mathrm{\phi\phi\phi}| \Bi \phi_\mathrm{s}(1-\phi_\mathrm{s})$, with $\Bi \phi_\mathrm{s}(1-\phi_\mathrm{s})/h_0$ the (dimensionless) rate of change of the volume fraction due to evaporation, as the two characteristic time scales that define the spinodal lag time~\cite{Schaefer2015StructuringEvaporation}
\begin{equation}\label{eq:early:tauL}
\tau_\mathrm{L} \approx 2^{5/3} r^{1/3} \left(\frac{\tau_\mathrm{d}}{\tau_\mathrm{e}^{2}}\right)^{1/3} \propto \Bi^{-2/3},
\end{equation}
and the emergent wave number $q_*$
\begin{equation}\label{eq:early:qstar}
    q_* \approx \left(\frac{1}{4\Cn} \frac{\tau_\mathrm{L}}{\tau_\mathrm{e}}\right)^{1/2} \approx \Cn^{-1/2} \left(\frac{r}{2} \frac{\tau_\mathrm{d}}{\tau_\mathrm{e}}\right)^{1/6} \propto \Bi^{1/6}.
\end{equation}
Here, $f_{\phi\phi\phi} = (1-\phi_\mathrm{s})^{-2} - \phi_\mathrm{s}^{-2}$ is the third derivative of the local free energy density Eq.~\eqref{eq:theory:FloryHuggins} evaluated at the low volume fraction spinodal, $r = \ln \delta \phi_{q_*}(\tau_\mathrm{L})/\delta \phi_{q_*} (0)$ is a measure for the amplification of the fluctuation amplitude that we associate with the spinodal lag time $\tau_\mathrm{L}$, which can in practice be treated as a fitting parameter, $M(\phi)$ the mobility defined in Eq.~\eqref{eq:theory:mobility}, and $\phi_\mathrm{s}$ the volume fraction at the low concentration spinodal~\cite{Schaefer2015StructuringEvaporation}. The factor $\phi_\mathrm{s}(1-\phi_\mathrm{s})/h_\mathrm{0}$ with $h_0$ the initial film height in the evaporation time scale $\tau_\mathrm{e}$ finds its origin in our solvent evaporation model. 
Our numerical calculations are in agreement with these predictions (not shown). Hence, we conclude that during early times the demixing kinetics in our quasi two-dimensional model is identical to those in a two-dimensional model without hydrodynamics, and depends non-trivially on diffusion and the rate of solvent evaporation that enter via two emergent time scales in Eqs.~\eqref{eq:early:tauL} and \eqref{eq:early:qstar}.

Next, we discuss the late-stage coarsening and how this is affected by the hydrodynamic transport, solvent evaporation and the coupling of the bulk with the fluid--gas interface.

\section{Late stage coarsening}\label{sec:coarsening}
As can be concluded from Fig.~\ref{fig:snapshots} and as discussed in more detail in the preceding Sections~\ref{sec:results} and \ref{sec:early}, the hydrodynamics of flow appears to mainly influence the morphology and associated characteristic feature size during the coarsening stage. While this is to be expected for critical or near-critical mixtures that show a bicontinuous demixed morphology~\cite{Siggia1979LateMixtures,Bray2002TheoryKinetics}, hydrodynamic coarsening for the typically off-critical dispersions of droplets that form in the context of our calculations is often believed to be of minor importance~\cite{Tanaka1996CoarseningMixtures}. This, of course, is not to say that hydrodynamic interactions between droplets do not play a role, e.g., in the compositional Marangoni effect associated with gradients in the solute-solvent interfacial tension~\cite{Shimizu2015AMixtures} or via the pumping action that coalescing droplets exert on the surrounding fluid~\cite{Tanaka1996CoarseningMixtures}, but these effects are relatively subtle. In this section, we show that in our model, hydrodynamics in combination with evaporation does have a strong impact on the coarsening under off-critical conditions in the sense that it speeds up the process in comparison to diffusive coarsening, starting at a time that decreases with increasing Peclet number. This resembles the effect of hydrodynamics on coarsening in bulk mixtures of critical composition, although the underlying mechanism turns out to be different~\cite{Bray2002TheoryKinetics}. By investigating how the coarsening dynamics depends on the Biot, Peclet and Capillary numbers, we are able to explain the origins of this kind of rapid coarsening. 

\begin{figure}
    \centering
    \includegraphics[width=0.75\columnwidth]{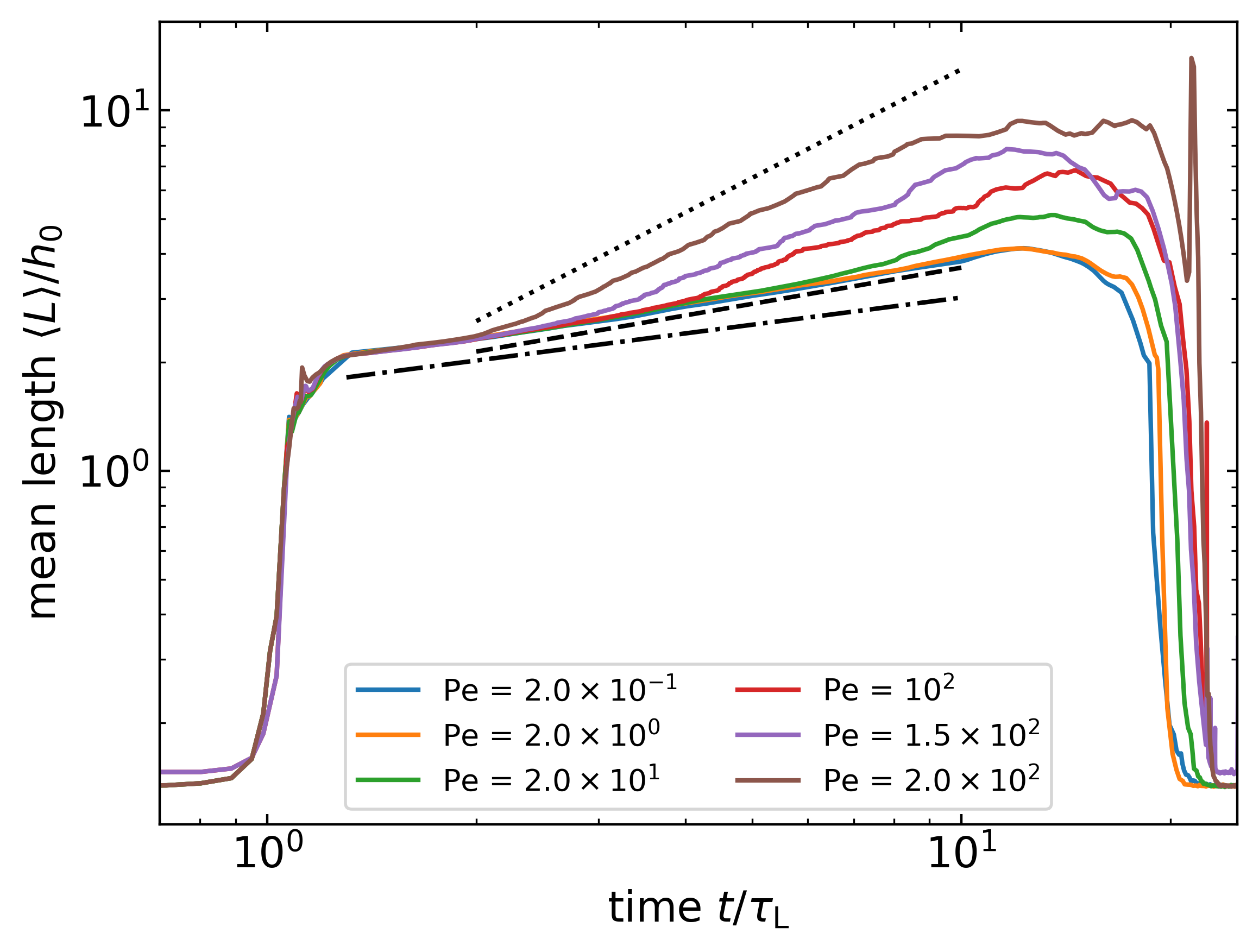}
    \caption{The mean compositional length scale $\langle L \rangle$ scaled to the initial film height $h_0$ as a function of the scaled time $t/\tau_\mathrm{L}$ for different values of the Peclet number between $\Pe = 2 \times 10^{-1}$ to $\Pe = 2 \times 10^{2}$ (bottom to top). The other model parameter values are $\Ca = 5\times 10^{-2}$, $\Bi = 3\times 10^{-3}$, $\chi = 4$, $G = 2.3 \times 10^{-4}$ and initial volume fraction $\phi_0 = 0.1464$. For $t/\tau_\mathrm{L} < 1$, the solution is still homogeneous and the mean length is of the order of the size of the spatial discretization, which is the length scale implicit in the \textit{discretized} thermal noise. For $t/\tau_\mathrm{L}>1$ the solution phase separates and subsequently coarsens. The dotted line indicates a scaling exponent of $\alpha = 1$, the dash-dotted lines $\alpha = 1/4$ and the dashed line $\alpha = 1/3$.}
    \label{fig:morphologicalLength}
\end{figure}

We characterize the coarsening kinetics by focusing attention on a characteristic compositional length scale $\langle L \rangle(t)$ that in the literature is generally assumed to obey the scaling relation $\langle L \rangle \propto t^{\alpha}$ with $\alpha$ an exponent. The value that this exponent takes depends on the predominant coarsening mechanism~\cite{Tanaka1996CoarseningMixtures,Mullins1986TheCoarsening}. Following standard procedure, we calculate the characteristic length from a mean characteristic wave number $\langle q \rangle(t)$, where $\langle L \rangle(t) = 2 \pi / \langle q \rangle(t)$, where $\langle q \rangle(t) \equiv \int \dint q q S(q)/\int \dint q S(q)$ and $S(q) = \langle |\delta \phi_q (t)|^2 \rangle$ the ensemble-averaged structure factor and  $\delta \phi_q (t)$ the Fourier transform of the fluctuation in the volume fraction defined in the previous section~\cite{Bray2002TheoryKinetics}. 
Fig.~\ref{fig:morphologicalLength} shows for fixed values of the Biot number $\Bi = 3\times 10^{-3}$ and the Capillary number $\Ca = 5 \times 10^{-2}$ for initial concentration $\phi_0 = \psi_0 = 0.1464$ the characteristic length $\langle L \rangle$ as a function of the scaled time for Peclet numbers ranging in value from $2 \times 10^{-1}$ to $2 \times 10^{2}$. Time is scaled to the spinodal lag time and the characteristic length scale to the initial height $h_0$ of the film. The ``spike'' in mean length just before the re-dissolution originates from the brief moment in time that only a single domain is present in our calculations and therefore is a finite-size effect.

What is immediately clear from the figure, is that late-stage coarsening strongly depends on the Peclet number, in particular for $\mathrm{Pe}\gg 1$.
As a guide to the eye, we have inserted a dotted line to indicate the scaling exponent $\alpha = 1$, a dash-dotted line for $\alpha = 1/4$ and a dashed line $\alpha = 1/3$. The solution demixes very swiftly switch at $t/\tau_\mathrm{L} = 1$, after which the characteristic length increases relatively slowly with time: the fluid film coarsens. For a while, the coarsening rate is an invariant of the Peclet number with a coarsening exponent close to albeit slightly larger than $\alpha = 1/4$. This is the expected coarsening exponent for a diffusive mobility that is of a ``double-degenerate'' form, \textit{i.e.}, large only in the solute-solvent interfaces, but (very) small in both solute and solvent-rich phases. The customary value of $\alpha = 1/3$ that Lifshitz-Slyosov-Wagner theory predicts holds only for constant or so-called one-sided mobilities~\cite{Lifshitz1961TheSolutions,Wagner1961TheorieOstwaldReifung,Dai2016ComputationalMobility}. 

For small Peclet numbers, the coarsening rate remains approximately constant during the coarsening stage until the morphology changes and reverses from a droplet phase with high solute concentration to one with relatively low solute concentration and subsequently redissolves. For Peclet numbers larger than about $\Pe = 20$, we find a transition of the coarsening exponent from $1/4$ to approximately $\alpha \approx 0.9$ -- we speculate that for larger Peclet numbers it actually approaches unity. The time of the transition shifts to earlier times with increasing Peclet number. Note that during this second coarsening regime the morphology remains that of a dispersion of droplets. Interestingly, the coarsening rate approaches that of viscous coarsening in three dimensions with $\alpha = 1$ albeit that viscous coarsening is only possible for bicontinuous and not for the droplet-like morphologies present in our calculations~\cite{Siggia1979LateMixtures,Tanaka1996CoarseningMixtures}. Even though the coarsening exponent becomes similar to that of viscous coarsening, the underlying mechanism turns out to be different. We return to this issue after discussing how the Biot and Capillary numbers impact the demixed morphology.

\begin{figure}
    \centering
    \includegraphics[width=0.75\columnwidth]{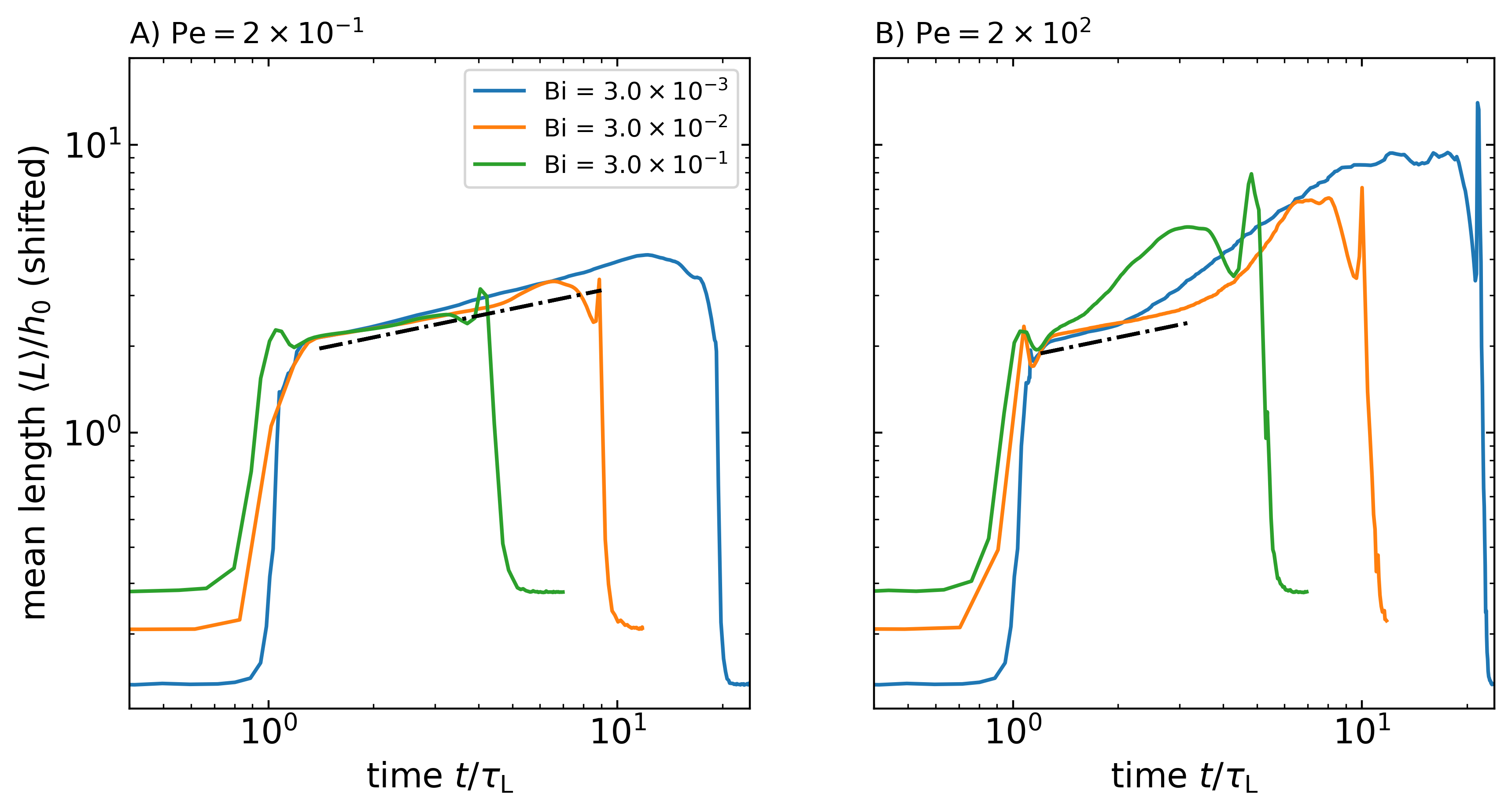}
    \caption{The (shifted) mean compositional length scale $\langle L \rangle$ scaled to the initial film height $h_0$ plotted as a function of the scaled time $t/\tau_\mathrm{L}$ for different values of the Biot number. The curves for $\Bi = 3\times 10^{-2}$ and $3\times 10^{-1}$ have been shifted vertically for the sake of comparability by a factor of, respectively, $1.6$ and $2.1$ such that they overlap at $t/\tau_\mathrm{L} \approx 1.2$. The model parameter values are $\Ca = 2\times 10^{-2}$, $\chi = 4$, $G = 2.3 \times 10^{-4}$ and initial volume fraction $\phi_0 = 0.1464$ for two different values of the Peclet number $\Pe = 2 \times 10^{-1}$ (A) and $\Pe = 2\times 10^{2}$ (B). The dash-dotted lines are a guide for the eye representing a coarsening exponent of $\alpha = 1/4$. See also the caption to Fig.~\ref{fig:morphologicalLength}.}
    \label{fig:Biot_Varies}
\end{figure}

In Fig.~\ref{fig:Biot_Varies}, we show for two values of the Peclet number $\Pe= 2\times10^{-1}$ and $\Pe = 2\times 10^{2}$ how different rates of solvent evaporation affect our results. The blue curves for $\Bi = 3 \times 10^{-3}$ are also shown in Fig.~\ref{fig:morphologicalLength}. To simplify comparison of the data for different Biot numbers, we shift the curves for $\Bi = 3 \times 10^{-2}$ and $\Bi = 3 \times 10^{-1}$ vertically, such that the curves overlap at $t/\tau_\mathrm{L} \approx 1.2$. Near $t/\tau_\mathrm{L} = 1$ we find that the mean length overshoots, an effect that appears to be more conspicuous at higher Biot numbers. This overshoot hints at the presence of a secondary length scale, which has already been observed and discussed by Schaefer and collaborators~\cite{Schaefer2015StructuringEvaporation} in the context of volatile solutions and we therefore do not discuss it here. Fig.~\ref{fig:Biot_Varies} shows that time available for coarsening decreases with increasing Biot number. Recall that $\tau_\mathrm{L} \propto \Bi^{-2/3}$ and that the drying time is proportional to $\Bi^{-1}$. Hence, the time available for coarsening differs by about a factor ten between the data, or in the units of scaled time $t/\tau_\mathrm{L}$ by a factor of $\Bi^{-1/3}$, so $10^{-1/3} \approx 0.46$. For $\Pe= 2\times10^{-1}$ shown in Fig.~\ref{fig:Biot_Varies}A, we again find the same coarsening exponent of about $1/4$, irrespective of the Biot number. 

For $\Pe = 2\times 10^{2}$, shown in Fig.~\ref{fig:Biot_Varies}B, increasing the Peclet number has a different effect depending on the value for the Biot number. For $\Bi = 3 \times 10^{-2}$, the coarsening exponent initially attains a value of about $0.2$, so below $1/4$, but since the scaling regime represents much less than a decade we should perhaps not read too much into this. The crossover to a power law of unity sets in subsequently, but again survives only for a fraction of a decade, after which coalescence of solute-rich droplets takes place, induced also by the decreasing distance between the solute-rich domains in response to the decreasing height of the film. For the fastest evaporation rate shown (green) the time available for coarsening is short and evaporation-induced material redistribution is faster than both diffusive or hydrodynamic transport. The coarsening exponent is approximately unity but applies again over a small period time before coalescence and re-dissolution take over. 
All in all, it seems that for small Peclet numbers, the Biot number has no significant impact other than to shorten the period in time over which coarsening can take place, at least if $\Bi<1$. This is not so for large Peclet number, in which case an increasing Biot number leads to a crossover to hydrodynamic behavior that depends non-monotonically on the Biot number. For $\Bi \gg 1$ and irrespective of the Peclet number the drying time eventually becomes shorter than the spinodal lag time $\tau_\mathrm{L}$, which prevents the solution from phase separating.

\begin{figure}
    \centering
    \includegraphics[width=0.75\columnwidth]{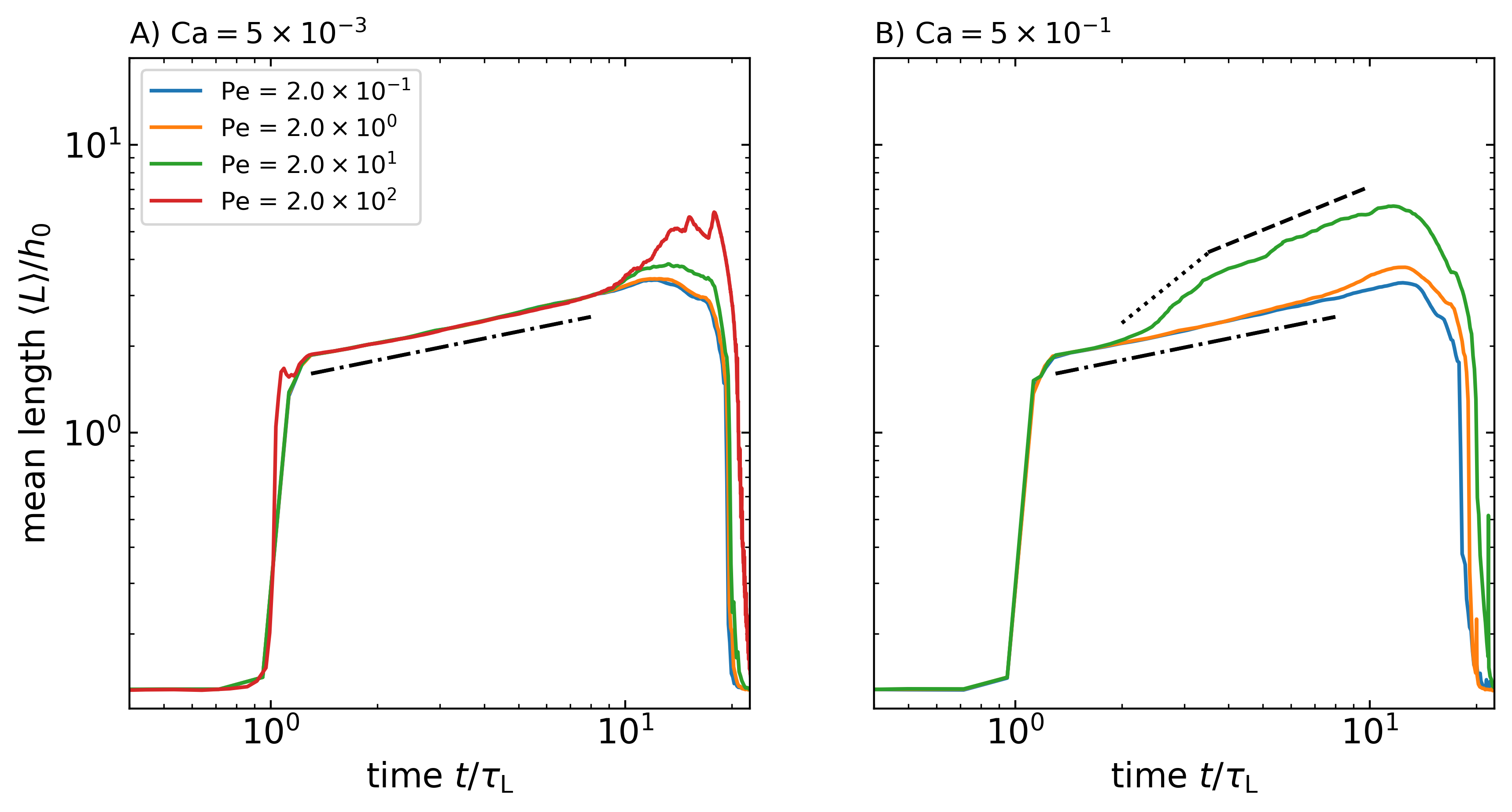}
    \caption{The mean compositional length scale $\langle L \rangle$ as a function of the scale time $t/\tau_\mathrm{L}$. The model parameter values are $\Bi = 3\times 10^{-3}$, $\chi = 4$, $G = 2.3 \times 10^{-4}$ and initial volume fraction $\phi_0 = 0.1464$ and (A) $\Ca = 2\times 10^{-1}$ and (B) $\Ca = 2\times 10^{-3}$, for the Peclet number $\Pe = 2 \times 10^{-1}$ in blue, $\Pe = 2 \times 10^{0}$ in orange and $\Pe = 2\times 10^{1}$ in green and $\Pe = 2\times 10^{2}$ in red. For $t/\tau_\mathrm{L}>1$ the solution phase separates and subsequently coarsens. The dash-dotted, dashed and dotted lines are guides for the eye for the coarsening exponents $\alpha = 1/4$, $\alpha = 1/2$ and $\alpha = 1$.}
    \label{fig:Capillary_Varies}
\end{figure}

Finally, in Fig.~\ref{fig:Capillary_Varies}, we show the coarsening rate for fixed value of the Biot number $\Bi = 3 \times 10^{-3}$ for two values of the Capillary number $\Ca = 5 \times 10^{-3}$ in Fig.~\ref{fig:Capillary_Varies}A and $\Ca = 5 \times 10^{-1}$ in Fig.~\ref{fig:Capillary_Varies}B, for Peclet numbers ranging between $2 \times 10^{-1}$ and $2 \times 10^{2}$. The dashed-dotted, dashed and dotted lines are guides for the eye to indicate coarsening exponents of $\alpha = 1/4$, of $\alpha = 1/2$ and of $\alpha = 1$. For very small Capillary number shown in Fig.~\ref{fig:Capillary_Varies}A, we find that for $t/\tau_\mathrm{L}< 10$ the coarsening is independent of the Peclet number with an exponent equal to approximately $\alpha = 1/4$ for about a decade in time, indicating that coarsening is diffusive. For a larger Capillary number, shown in Fig.~\ref{fig:Capillary_Varies}B, we obtain what resembles diffusive coarsening with $\alpha \approx 1/4$ for $\Pe < 2 \times 10^{1}$. For $\Pe = 2 \times 10^{1}$, we find a transition in the coarsening rate similar to what we found earlier for $\Ca = 5 \times 10^{-2}$ in Fig.~\ref{fig:morphologicalLength}. However, there seems to be a second transition to a slower coarsening corresponding to an exponent of $\alpha \approx 1/2$ albeit that we do not quite understand the physics underlying this transition nor that of the slower rate of coarsening. The data for $\Ca=0.5$ and $\Pe = 2 \times 10^{2}$ are not shown in Fig.~\ref{fig:Capillary_Varies}B, as in this case hydrodynamic transport already becomes important during the demixing stage, resulting in a much more rapid increase in the characteristic feature size and finite-size effects are large, preventing us from interpreting the results.

Basing ourselves on the results shown in Figs.~\ref{fig:morphologicalLength}, \ref{fig:Biot_Varies} and \ref{fig:Capillary_Varies}, we conclude that the transition from diffusive to any of the faster coarsening modes depends on the predominance of hydrodynamic transport (described by the Peclet number) and that of the fluid-gas surface tension relative to fluid-fluid interfacial tension (described by the Capillary number). While we obtain similar coarsening rates for the non-diffusive coarsening mode for low and high Biot number, we actually identify two distinct coarsening mechanisms. For high Biot numbers, the drying time is very short and so the available time for coarsening is short also. At high Biot and high Peclet number, the inter-droplet distance decreases rapidly, which facilitates the merging of the solute-rich domains, a process aided by hydrodynamic interactions. 

For small Biot and large Peclet numbers, we discover in Fig.~\ref{fig:morphologicalLength} and Fig.~\ref{fig:Capillary_Varies}B a similar transition in the coarsening rate, with a coarsening exponent changing from $\alpha = 1/4$ to about $\alpha = 0.9$. We associate this with a different and as far as we are aware, novel coarsening mechanism that we refer to as confluent coarsening. Since this mechanism only emerges at sufficiently high Peclet and Capillary numbers, we argue that it is related to the hydrodynamic transport processes originating from the (curved) solution-gas and the solute-solvent interfaces. In the next section, we focus attention on the flow fields and associated hydrodynamic transport mechanisms to unveil the physical origins of confluent coarsening. We find that at its root is the coupling of hydrodynamic transport in the phase-separating solution to gradients in the height of the liquid-gas interface. This coupling results in the directional motion of solute-rich droplets that accumulate in regions of the film where the film is relatively thin. These domains subsequently coalesce, resulting in enhanced domain growth. 

\section{Flow fields and transport mechanisms}\label{sec:flowfields}
The rapid coarsening that we find for the combination of sufficiently large Peclet and Capillary numbers in Figs.~\ref{fig:morphologicalLength} and \ref{fig:Capillary_Varies}B, indicates that hydrodynamic transport can have a strong influence on the late-stage morphology. In stark contrast with bulk mixtures, this is true even for off-critical mixtures as Fig.~\ref{fig:snapshots} also illustrates. In this section, we analyze the hydrodynamic transport processes by visualizing the flow fields that we calculate using Eq.~\eqref{eq:theory:velocityfield}. From our analysis, we find that the solute-rich droplets tend to move advectively and that the droplet motion aligns with gradients in the height of the film. We explain why this motion is appreciable only if the Peclet and Capillary numbers are sufficiently high, or, equivalently if (i) hydrodynamic transport is sufficiently rapid and (ii) the solution-vapor surface is (relatively) easily deformed at the three phase contact lines. The regions where the film is relatively thin act in some sense as focal points for the droplets to accumulate and coalesce. This increases the rate of growth of solute-rich domains and therefore results in enhanced coarsening. 

\begin{figure}
    \centering
    \includegraphics[width=\columnwidth]{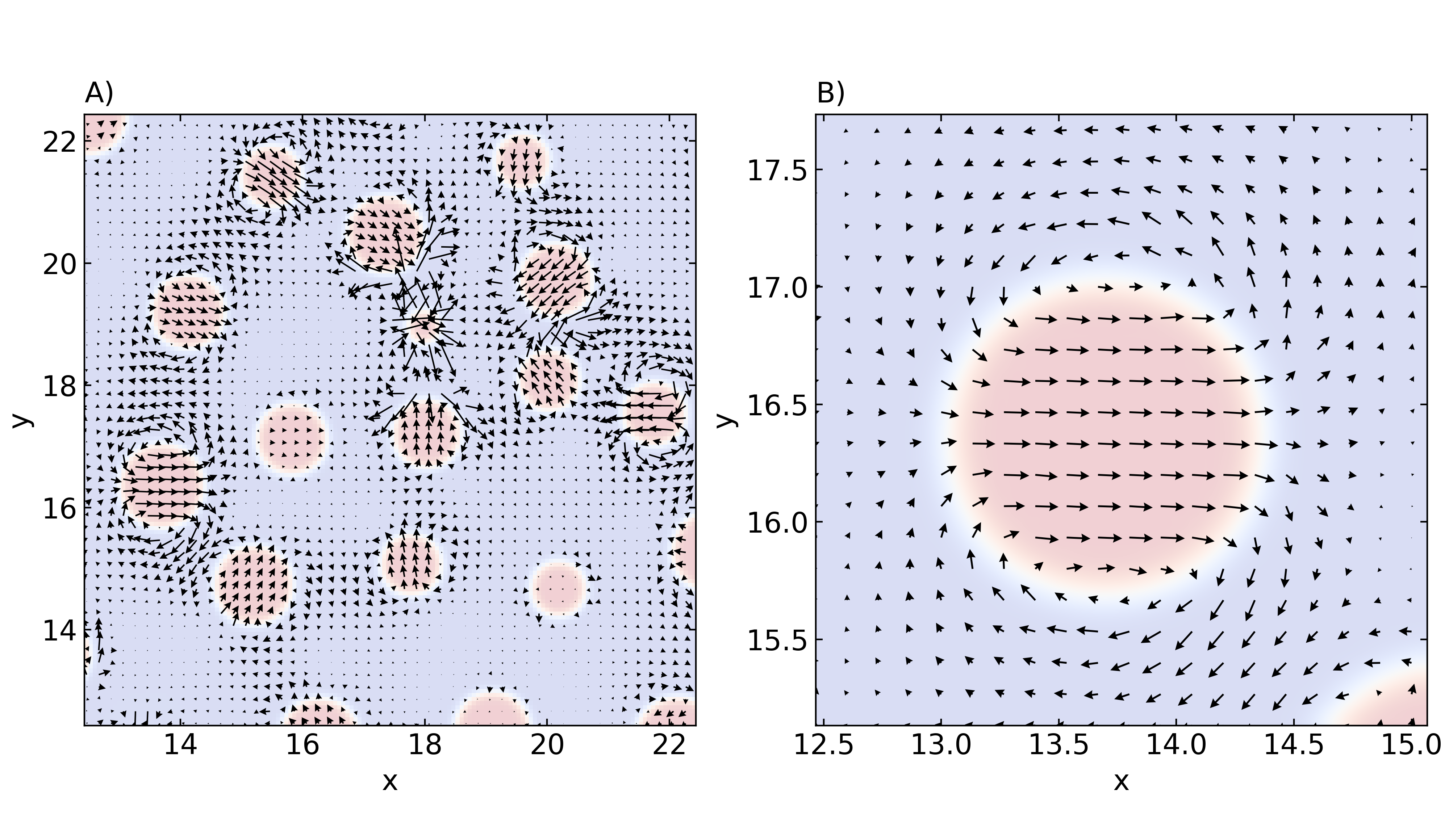}
    \caption{Rendering of the quasi two-dimensional flow field in the late stages of demixing of our binary fluid. The black arrows indicate the direction and magnitude of the fluid velocity field in the laboratory frame for $\Pe = 2\times10^{2}$, $\Ca = 5 \times 10^{-2}$, $\Bi = 3\times 10^{-3}$ and $t/\tau_\mathrm{L} = 3.15$, equivalent to the bottom left snapshot of Figs.~\ref{fig:snapshots}B and D. For clarity, we do not indicate local fluid velocities smaller than 1\% of the maximum fluid velocity. Panel A: overview. Panel B: enlarged flow field around one of the droplets of A. In both panels, we superimpose the fluid-velocity field with the volume fraction field $\phi$ in red indicating the solute-rich phase and in blue the solute-poor phase. The domains translate from left to right. }
    \label{fig:velocityfields_Solo}
\end{figure}

To highlight the directional motion of the solute-rich droplets, we show the velocity field in the laboratory frame in Fig.~\ref{fig:velocityfields_Solo} for $\Pe = 2 \times 10^{2}$, $\Bi = 3 \times 10^{-3}$ and $\Ca = 5 \times 10^{-2}$ for $t/\tau_\mathrm{L} = 3.15$. The corresponding compositional snapshot is shown in the bottom left panel of Fig.~\ref{fig:snapshots}C. We superimpose the velocity field on the local solute concentration field, where high solute concentration is colored red and low solute concentration is colored blue. As Fig.~\ref{fig:velocityfields_Solo}A and B show, where in the latter we zoom in on a single solute-rich domain, the fluid velocity field inside the solute-rich domains is approximately uniform in both direction and magnitude. See also Fig.~9 in the supplemental material for a comparison of the fluid velocity of the droplet shown in Fig.~\ref{fig:velocityfields_Solo}B in the laboratory and centre-of-mass reference frame. The direction of motion of the droplets suggest that the droplets move towards a common region in the domain shown, the reason for which we explain below. In the solvent-rich phase, indicated in blue, the velocity field circles around the solute-rich domains, highlighted in the closeup image of Fig.~\ref{fig:velocityfields_Solo}B. It shows that the droplet pushes away solvent on the right side of the droplet and that this solvent is transported to the wake of the droplet, on the left side of it in Fig.~\ref{fig:velocityfields_Solo}B. At the phase boundaries perpendicular to the direction of motion of the droplet, vortices can be seen with a clockwise and counter clockwise direction, reminiscent of a vortex dipole.

\begin{figure}
    \centering
    \includegraphics[width=\columnwidth]{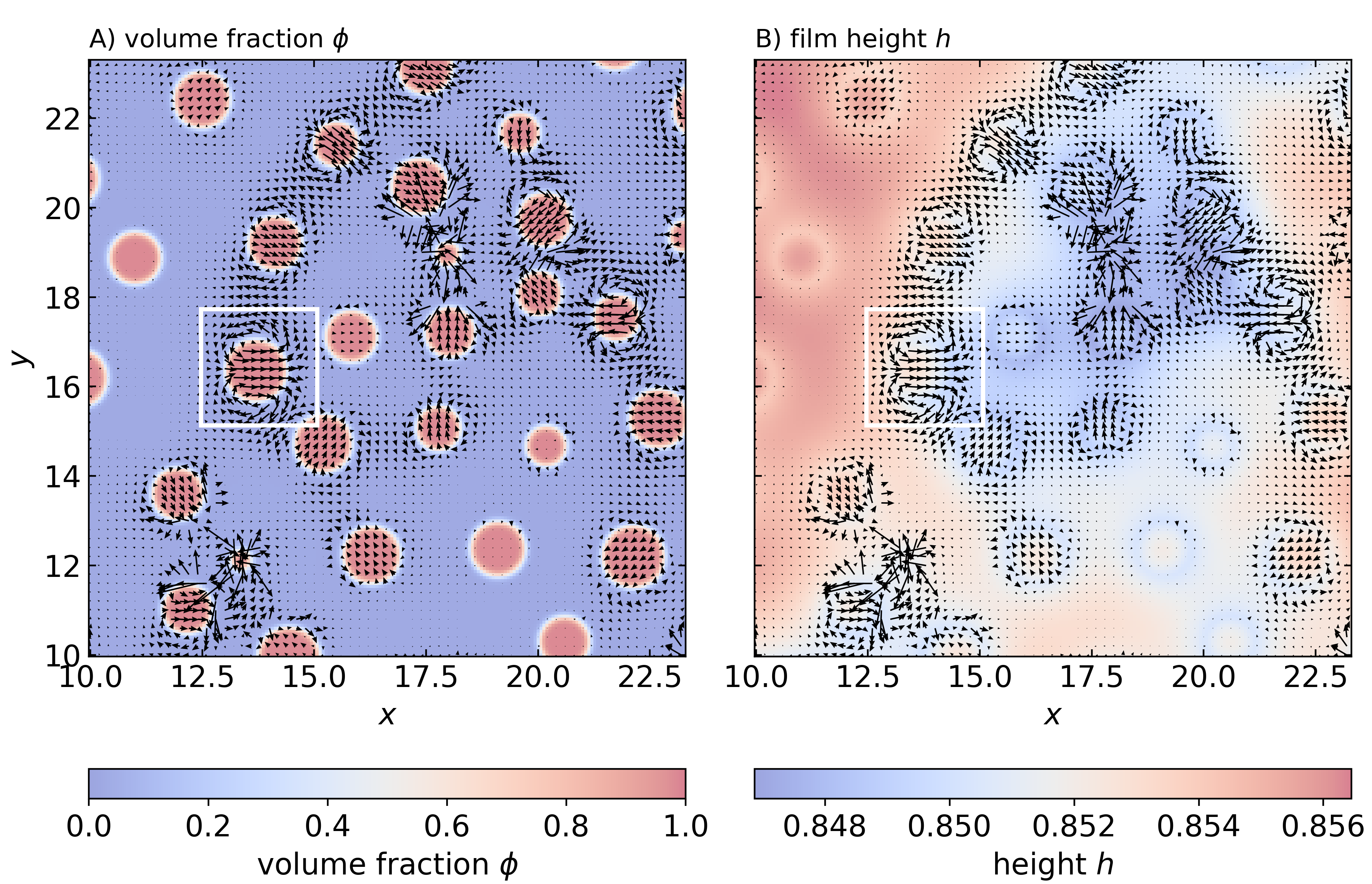}
    \caption{The quasi two-dimensional flow field in the laboratory frame (black arrows) for $\Pe = 2\times10^{2}$, $\Ca = 5 \times 10^{-2}$, $\Bi = 3\times 10^{-3}$ at $t/\tau_\mathrm{L} = 3.15$, equivalent to the bottom left panel of Figs.~\ref{fig:snapshots}B and D. For clarity, we do not not show the local fluid velocities smaller than 1\% of the maximum fluid velocity. In panel A, we superimpose the fluid-velocity field with the volume fraction field $\phi$, red representing the solute-rich phase and blue the solute-poor phase. In panel B, we superimpose the fluid-velocity field with the height field $h$, red indicating relatively high regions and blue relatively low-lying regions. The white box indicates the enlarged flow field shown in Fig.~\ref{fig:velocityfields_Solo}B.}
    \label{fig:velocityfields_Duo}
\end{figure}

While the fluid velocity shown in Fig.~\ref{fig:velocityfields_Solo} correlates with the presence of solute-rich domains, the direction of motion does not, that is, there is no discernible gradient in the concentration of solute that correlates with it. Instead, it is correlated with the \textit{slope} of the height of the liquid-gas interface. This we show in Fig.~\ref{fig:velocityfields_Duo}, presenting in \ref{fig:velocityfields_Duo}A the fluid velocity field superimposed on the concentration field and in \ref{fig:velocityfields_Duo}B the height of the corresponding solution-gas interface. The volume fraction and height color bars are shown below the figures. As we deduce from Fig.~\ref{fig:velocityfields_Duo}B, the direction of motion of the solute-rich droplets clearly aligns with the gradients in the height of the film. The domains move deterministically from regions where the film is thick to regions of space where the film is thin. Hence, regions where the film is thin appear to represent areas where droplets accumulate. This facilitates the coalescence of droplets, which eventually leads to an increase in the average domain size. 

We need to explain three things: (i) why the droplet motion couples to gradients in the height of the film, (ii) why gradients in the height of the film emerge in the first place and (iii) how our model parameters affect droplet motion. The latter we explain while answering the first two questions. To answer the first question, we draw the attention of the reader to the expression for the velocity field in Eq.~\eqref{eq:theory:velocityfield}. Only the last term in Eq.~\eqref{eq:theory:velocityfield} that in dimensionless units reads $-\Pe \sqrt{\Cn} h (\nabla h \cdot \nabla \phi) \nabla \phi/3$ couples the bulk hydrodynamics of the liquid-liquid phase boundaries to gradients in the height of the film. With this in mind, let us take the single droplet highlighted with the white box in Fig.~\ref{fig:velocityfields_Duo}, which is the same droplet as shown in Fig.~\ref{fig:velocityfields_Solo}B, as an example to investigate how precisely this contribution affects the fluid motion. In the region within the white box the height of the film decreases with increasing $x$-coordinate, and the motion of the droplet is also in that direction. We assume for our argument that the droplet shown is perfectly circular. We only need to focus on the fluid-fluid phase boundaries, because $\nabla \phi$ is negligibly small outside of these phase boundaries. For the fluid-fluid interface on the left hand side of the center of mass the term $(\nabla h \cdot \nabla \phi)$ is negative and $\nabla \phi$ is positive, whereas for phase boundaries on the other side $(\nabla h \cdot \nabla \phi)$ and $\nabla \phi$ are both negative. Hence, we expect the droplet highlighted within the white box in Fig.~\ref{fig:velocityfields_Duo} to move from left to right, which is indeed what we observe. Since the magnitude of the velocity is proportional to the Peclet number, this also explains why this motion and therefore confluent coarsening is only noticeable for sufficiently large Peclet numbers. 

What remains is an explanation for the origin of (long-ranged) gradients in the height of the film. These gradients do not originate from the three-phase contact line of a single droplet because this results in a relatively short-ranged and isotropic deformation of the solution-gas surface. Instead, the gradients in the height of the film form during the initial stages of demixing. While the demixed morphology becomes that of solute-rich droplets dispersed isotropically in a solvent-rich majority phase, our numerical calculations indicate that the kinetics of the initial demixing process is not spatially homogeneous: the solute-rich domains tend to emerge somewhat clustered. Hence, for a very brief period of time the phase-separated morphology is that of a collection of clustered solute-rich domains, while some regions in the film have not yet fully phase separated. The downward force exerted on the solution-gas interface by these clusters of domains then causes a collective, larger scale deformation of the film surface. These deformations persist even after the solution is phase-separated everywhere in the film, resulting in the gradients in the height of the film required for the directional motion of the droplets. This process turns out to be regulated by the Capillary number $\Ca$.

To show that this must be so, we take as characteristic measure of the slope $\nabla h = \Delta h/\Delta L$, where $\Delta h$ is the magnitude of the deformation of the height of the film and $\Delta L$ the typical length scale of the deformation. We are able to get an estimate of $\Delta L$ from the Laplace pressure $\Delta P = \Delta F/A = \gamma/ \Delta L$, with $\Delta F$ the force exerted by a cluster of solute-rich domains on the solution-gas interface, $A$ the area over which the force is exerted and $\gamma$ the solution-gas surface tension. (See also Eq.~\ref{eq:theory:free_energy}.) Hence $\Delta L \propto \gamma \propto \Ca^{-1}$. For $\Delta h$ we assume that the solution-gas interface has Hookean elasticity, suggesting $\Delta h \propto 1/\gamma\propto \Ca$. We deduce that $\nabla h \propto \Delta h/\Delta L\propto \Ca^2$. 
All of this suggests that we can estimate the droplet velocity as $\mathbf{u} = -\Pe \sqrt{\Cn} \hspace{0.1cm} h (\nabla h \cdot \nabla \phi) \nabla \phi/3 \propto \Pe \hspace{0.1cm} \Ca^{2}/\Cn^{1/2}$, using the fact that $\sqrt{\Cn}$ is a measure for the width of the liquid-liquid interface, and therefore that $\nabla \phi \propto \Cn^{-1/2}$. Hence, we find that the relevant dimensionless parameters that set the droplet motion are the Peclet, Capillary and Cahn numbers. For confluent coarsening to be dominant, the droplet velocity must be sufficiently high for the motion to be perceivable within the time window of our numerical experiment. In other words, if the Peclet and Capillary number are sufficiently large and the Cahn number sufficiently small. We now expect a transition from diffusive to confluent coarsening if the droplets have translated a sufficiently large distance, hence the time of the transition must be inversely proportional to the droplet velocity; it decreases with increasing Peclet, Capillary and increases with decreasing Cahn number (the latter we have not verified). This is in agreement with our findings presented in Figs.~\ref{fig:morphologicalLength} and \ref{fig:Capillary_Varies}.

While this discussion explains the origin of confluent coarsening and how it depends on the parameters of our model, we have not attempted to theoretically predict the value for the coarsening exponent that is associated with it. Further, in the light of the above discussion where the Biot number does not play a role, we expect that confluent coarsening should occur irrespective of the solvent evaporation rate, at least if the drying time is not shorter than the typical time required for the droplets to move a sufficient distance. Indeed, we find by means of calculations on non-volatile off-critical mixtures (not shown) that the kind of directional transport required for confluent coarsening is present also. Interestingly, we do \textit{not} observe this coarsening mechanism in our calculations on non-volatile mixtures of critical composition (not shown), even though similar longer-ranged gradients in the height of the film are present in the calculations. Hence, we conclude that for non-volatile (near-)critical mixtures the morphology ripens via another hydrodynamic coarsening mechanism, \textit{i.e.}, viscous coarsening, which suppresses confluent coarsening. In other words, confluent coarsening appears to emerge only for off-critical mixtures.

\section{Discussion and conclusion}\label{sec:discussion_conclusion}
In summary, we have theoretically studied the evaporation-driven phase separation of an incompressible binary fluid in a thin film. It is a model for the fabrication of solution-processed thin films in which typically the deposition of the film on a solid support is so fast that phase separation occurs far removed from the deposition apparatus and associated capillary zone, \textit{i.e.}, the zone close to the deposition apparatus where mass transport is dictated by the deposition technique itself, e.g., in meniscus-guided deposition. Away from the capillary zone the film is for all intents and purposes flat, so no longer influenced by the curvature of the film in the capillary zone, and the fluid velocity uniform and equal to that of the substrate. This means that the problem reduces to that of a flat, stationary film.

In our model, the solution is bounded by a non-deformable, flat and neutral substrate and by a free interface with the gas phase. The solvent is volatile and evaporates with a rate that is proportional to its volume fraction. We focus on conditions where vertical stratification induced by the evaporation or phase-separation cannot occur, and allow for both diffusive and advective mass transport within the so-called lubrication approximation. The three \textit{main} dimensionless groups in our model are the Peclet number, the Capillary number and the Biot number. The first describes the importance of hydrodynamic transport of material relative to diffusive transport, the second measures the relative strength of the liquid-gas and the liquid-liquid interfacial tensions and the third expresses the strength of evaporation relative to diffusion. We define two additional dimensionless groups, being the disjoining number and the Cahn number, which describe the strength of the liquid-liquid capillary forces relative to the van der Waals forces and the width of the liquid-liquid interfaces. In our calculations we keep the magnitude of these two dimensionless numbers fixed.

The demixing of the solution tends to occur under off-critical conditions, which is a result of solvent evaporation gradually destabilizing the solution starting from a solute concentration equal to the low concentration branch of the spinodal. Hence, irrespective of the values of the dimensionless groups, the morphology initially is that of solute-rich droplets dispersed in a solvent-rich majority phase. The morphology eventually reverses to that of solvent-rich droplets in a solute-rich majority phase and subsequently redissolves due to ongoing evaporation. Associated with the compositional morphology is structure formation in the height of the film. This structure emerges due to disparity in the rate of evaporation in the regions rich in either solute or solvent and the downward force exerted on it at the three-phase solute-solvent-gas contact lines. The resulting roughness of the free surface affects the drying kinetics, which we find to be more rapid at a higher degree of surface roughness, \textit{i.e.}, for small Peclet and large capillary numbers. 

During the early stages of demixing the temporal evolution of the height and volume fraction fields are, in principle, coupled on account of solvent evaporation, irrespective of the values for the Peclet and Capillary numbers. Mimicking the setup of our numerical calculations, wherein (thermal) fluctuations in the height are only excited indirectly via the thermal fluctuations in the volume fractions, we argue that in that case this coupling is weak and can actually be neglected. The initial stages of demixing are therefore dictated by diffusion and solvent evaporation only, and we reproduce the work of Schaefer \textit{et al.} that assumes a perfectly flat solution-gas interface and ignores the presence of hydrodynamic flow fields altogether~\cite{Schaefer2015StructuringEvaporation,Schaefer2016StructuringEvaporation}. The relevant dimensionless groups that set the kinetics of the early stages of demixing are the Cahn and Biot numbers. These results seemingly need to be modified for the case where thermal fluctuations in the height of the film are excited directly. Interestingly, this appears to be true only for volatile mixtures because the coupling of bulk and surface fluctuation modes appears to be mediated via solvent evaporation, whereas for non-volatile mixtures the bulk and surface modes remain decoupled. How exactly these fluctuations affect the early stages of demixing we leave for future work.

In the late stages of demixing, we discern a number of different coarsening modes as summarized in Fig.~\ref{fig:4_Schematic}. Each mode is associated with one or more of the transport mechanisms at play in our model calculations and accompanied by a different coarsening exponent. Apart from the well-known Ostwald-type ripening with in our case a coarsening exponent of one-fourth and an earlier predicted evaporative coarsening regime~\cite{Schaefer2015StructuringEvaporation,Negi2018SimulatingInvestigation}, we find that hydrodynamics has a strong impact on the coarsening behavior for high Peclet and Capillary numbers. For volatile mixtures that phase separate under off-critical conditions, we identify two coarsening modes that originate from the interplay between different hydrodynamic effects. The first only emerges for fast evaporation and high Peclet numbers. Here, the balance between evaporation and Laplace-pressure-driven material redistribution rapidly decreases the inter-droplet distance, which in turn promotes the coalescence of domains and hence the coarsening process. The coarsening exponent appears somewhat larger than unity, but only persists for a short period of time and we therefore cannot accurately determine it. The second coarsening mode emerges for high Peclet and Capillary numbers, so if the ratio of the liquid-liquid and liquid-gas interfacial tensions is sufficiently large. 

This second mode is, as far as we are aware, a novel coarsening mechanism, which we find to be present only if the morphology is a dispersion of droplet-like phase-separated domains, rather than bicontinuous. We refer to this coarsening mechanism as confluent coarsening. It finds its origin in the interplay of the three phase liquid-liquid-gas contact lines with the gradients in the thickness of the liquid film. The gradient in the film height emerges during the initial stages of demixing. By analyzing the fluid flow fields, we show that this results in directional motion of these domains towards regions of space where the film is relatively thin, which facilitates the coalescence of droplets and results in fast coarsening. Indeed, we find a coarsening rate with an exponent of approximately unity. Interestingly, this coarsening exponent is similar to that observed for a three-dimensional bicontinuous bulk morphology that evolves via viscous coarsening albeit that this latter process is governed by a completely different mechanism. We do not provide a theoretical explanation for the observed coarsening rate, which we also leave for future work.

As far as we are aware, confluent coarsening has not yet been identified experimentally. The reason for this may be that it is suppressed by a potentially strong increase in the viscosity of the solution caused by solvent evaporation and demixing, suppressing advective transport in the film. It might also be that other mechanisms, not part of our model description, become important, such as Marangoni effects originating from gradients in the liquid-gas surface tension, caused by either gradients in the composition or the temperature. In future work, we aim to address the limitations in our model by including thermal and solutal marangoni effects, as well as a concentration dependent viscosity. Moreover, a comparable study wherein we relax the lubrication approximation may be required in the presence of strong Marangoni fluxes~\cite{Oron1997Long-scaleFilms,Naraigh2010NonlinearFilms}.

\section{Supplemental Material}

\begin{figure}[h]
    \centering
    \includegraphics[width=\columnwidth]{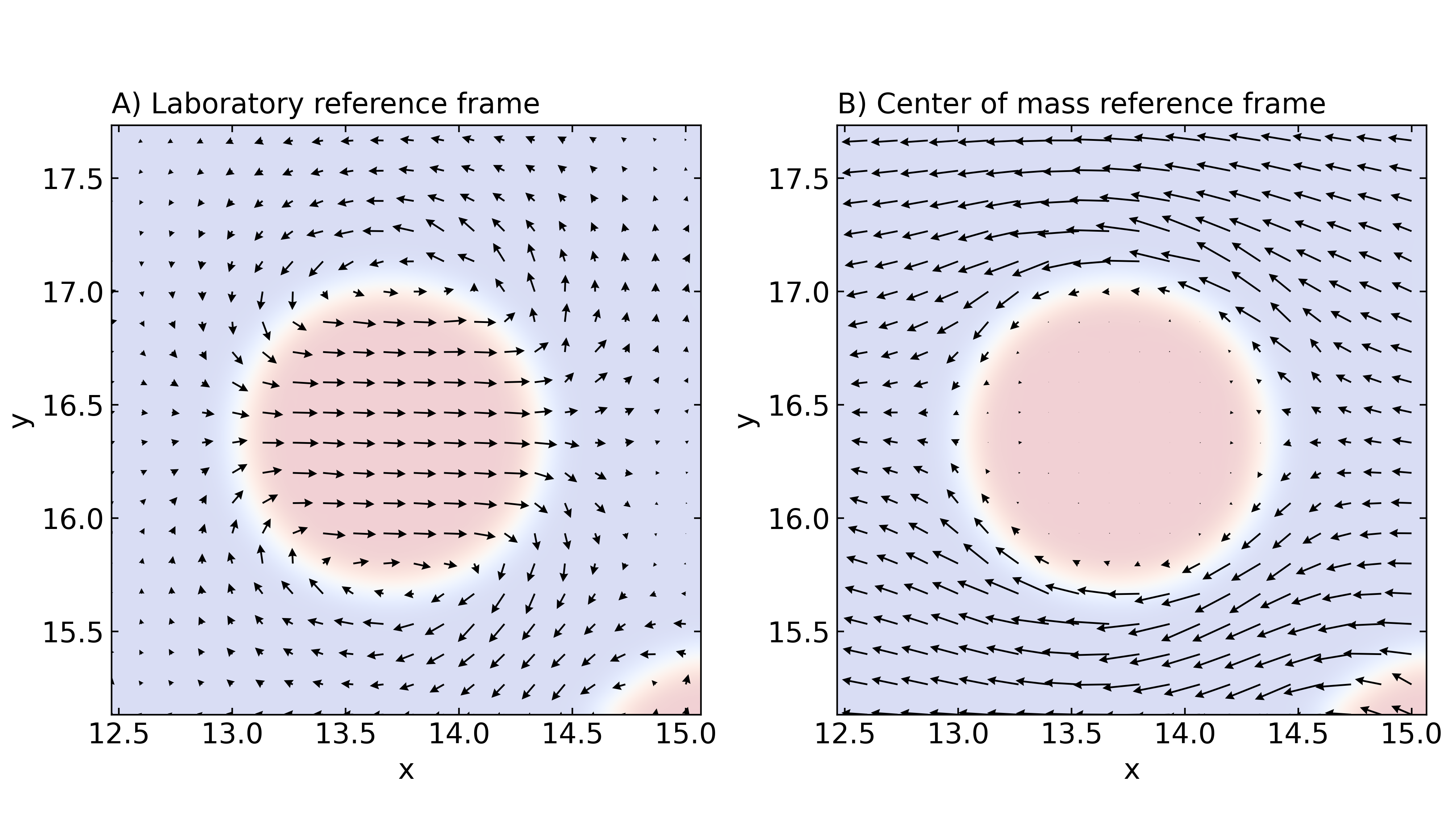}
    \caption{Rendering of the quasi two-dimensional flow field in the late stages of demixing of our binary fluid around a single solute-rich droplet. The black arrows indicate the direction and magnitude of the fluid velocity field. We superimpose the fluid-velocity field with the volume fraction field $\phi$ in red indicating the solute-rich phase and in blue the solute-poor phase. Panel A is identical to Fig.~7B, and the black arrows indicate the direction and magnitude of the fluid velocity field in the laboratory frame. In panel B the black arrows indicate the flow field in the centre-of-mass frame of the droplet. The parameter values are $\Pe = 2\times10^{2}$, $\Ca = 5 \times 10^{-2}$, $\Bi = 3\times 10^{-3}$ and $t/\tau_\mathrm{L} = 3.15$, equivalent to the bottom left snapshot of Figs.~2B and 2D. For clarity, we do not indicate local fluid velocities smaller than 1\% of the maximum fluid velocity.}
    \label{fig:velocityfields_Lab_COM}
\end{figure}

\bibliography{references}

\end{document}